\documentclass[%
 aip,
 jcp,
 amsmath,amssymb,
 reprint,%
]{revtex4-1}

\usepackage{graphicx}
\usepackage{dcolumn}
\usepackage{bm}

\usepackage[utf8]{inputenc}
\usepackage[T1]{fontenc}
\usepackage{mathptmx}
\usepackage{etoolbox}
\usepackage{xcolor}
\usepackage{placeins}

\newcommand{\MXEL}[3]{\left< #1 | #2 | #3 \right>}
\newcommand{\KET}[1]{\left\vert #1 \right>}
\newcommand{\RE}[1]{\textrm{Re}\,#1\,}
\newcommand{\IM}[1]{\textrm{Im}\,#1\,}
\newcommand{\BRAKET}[2]{\left< #1 \vert #2 \right>}
\usepackage{enumitem}
\usepackage{xcolor}

\makeatletter
\def\@email#1#2{%
 \endgroup
 \patchcmd{\titleblock@produce}
  {\frontmatter@RRAPformat}
  {\frontmatter@RRAPformat{\produce@RRAP{*#1\href{mailto:#2}{#2}}}\frontmatter@RRAPformat}
  {}{}
}%
\makeatother
\begin{document}

\preprint{AIP/123-QED}

\title[QUAPI with non-commuting baths]{Numerically efficient quasi-adiabatic propagator path integral approach with two independent non-commuting baths}
\author{R.~Ovcharenko}
\author{B.~P.~Fingerhut}%
 \email{roman.ovcharenko@cup.lmu.de, benjamin.fingerhut@cup.lmu.de}
\affiliation{ 
Department of Chemistry and Centre for NanoScience, Ludwig-Maximilians-Universität München, 81377 München, Germany.
}%

\date{\today}

\begin{abstract}
Path integral methods, like the quasi-adiabatic propagator path integral (QUAPI), are widely used in general-purpose and highly accurate  numerical benchmark simulations of open quantum systems, particularly in regimes inaccessible to perturbative methods.
Nevertheless, the applicability of the QUAPI method to realistic systems of interest is restricted by the exponentially growing computer memory requirements with respect to the size of the quantum system and the time range of non-Markovian correlation 
effects. This exponential ``wall'' becomes even more severe for multiple non-commuting fluctuating environments.
In the present work, we address the numerical efficiency and accuracy of approximations that have been introduced for the QUAPI method with a single general environment, for the case of two independent non-commuting environments where one of them is considered as a pure dephasing environment.
Specifically, we consider a sharply defined cut-off of the memory time,  path filtering and mask assisted coarse 
graining of influence functional coefficients (MACGIC-QUAPI) as approximations. We demonstrate that commonly applied numerical techniques such as path filtering cannot be straightforwardly transferred to the two bath case even in the weak-coupling and quasi-Markovian limits. On the other hand, the sharply defined memory cut-off can be accurately handled with the mask assisted coarse 
graining (MACGIC-QUAPI) approach.
Our findings demonstrates that if system coupling operators to different baths do not commute, the additive nature of the statistically independent environments may be misleading.
Particularly, the quasi-Markovian nature of a pure dephasing 
bath is lost, once there simultaneously exists another non-commuting source of fluctuations. 
%
\end{abstract}

\maketitle

%

\section{Introduction\label{sec:intro}}
The inevitable nature of the environment and its destructive influence on the coherence properties of quantum systems led to the necessity of the theoretical descriptions of the dynamics of open quantum systems~\cite{a:Redfield:57, a:Feynman:63, a:Leggett:87}.
Dissipative effects, like population relaxation and decoherence are the consequences of the interaction of a system with various sources of fluctuations.
Often, dissipation induced by an environment plays a crucial role for a vast range of complex physical phenomena, such as excitation energy or charge transfer in solids~\cite{a:Topaler:93, a:Golding:92, a:Vojta:01} or 
biological photosynthetic complexes~\cite{b:Amerongen:00,b:May:11,a:Ishizaki:09, a:Ishizaki:09b,a:Nalbach:10, a:Lambert:12,a:Richter:17,a:Richter:19}.
For example, the interaction of an electronic system with nuclear modes of the environment can induce complex vibronic relaxation dynamics. Particularly intriguing is the example of  anti-correlated vibrations in resonance with the electronic system that acts as driver of energy transfer.\cite{Tiwari:PNAS:13} Such scenario can impact the beating pattern and line shapes of multi-dimensional spectra\cite{Hybl:CPL:98,Jonas:AnnRevPhysChem:03} in a non-trivial way, calling for robust, non-perturbative simulation methods.
On the other hand, the environment may represent a destructive noise source, effectively spoiling desired microscopic properties of a system prepared in a specific quantum state. 
The latter representation of the environment as noise recently received major attention due the progress in the development of quantum computers~\cite{a:Shor:95,Gambetta:NPJ:07}. 

Qubits, as elementary unit of quantum computers and despite the particular realization~\cite{a:Nielsen:10, a:Lund:08,a:Monz:09,a:Neumann:10,a:Song:17,a:Mourik:12,a:Morton:08} are two-level quantum systems (TLS) that are subject to environmental fluctuations.
For simplicity, these bath fluctuations are often grouped together to form a single effective bath. 
In this view, a qubit immersed into its environment is reduced to the spin-boson 
model~\cite{a:Leggett:87} that has been intensively investigated over past few decades~\cite{a:Leggett:87,Vega:RevModPhys:17}. 
A concept of a single effective bath implies a weak coupling of all fluctuating environments to the quantum system so that the total influence of all 
baths is cumulative. In practice, however, such additive properties are not always satisfied (see Ref.~\citenum{a:Yao:15} and references therein). Different baths 
interfere with each other, constructively or destructively, that may decrease the resulting decoherence or relaxation rates. Moreover, experimentally, 
dephasing is typically a much faster process than relaxation, that suggests more dephasing channels to exist than relaxation ones. This give rise to an idea 
that more than one effective bath is necessary for a quantitative understanding. 

Some experimental qubit realizations are even designed in such a way that suggests at 
least two baths of different nature. Realizations of quantum dot based qubits \cite{a:Allen:12,a:Karzig:17} and superconducting qubits~\cite{Blais:PhysRevA:04,Gambetta:NPJ:07}
are prototypical examples. In both, a quantum TLS is affected by two independent voltage gates treated as 
fluctuations. Acting on a TLS, one source of fluctuations affects the system energy gap, while the other changes the tunneling barrier. To theoretically describe such settings, one may extend the spin-boson model to contain two independent baths coupled to the TLS via the $\sigma_z$ and $\sigma_x$ Pauli matrices, respectively. 

Theoretical approaches for open quantum systems range from 
second-order perturbation theory master equation approaches, such as the Redfield~\cite{a:Redfield:57} and 
Lindblad~\cite{a:Lindblad:76} equation, to higher order methods, like the resumed perturbative theories (RESPET),~\cite{a:Wuerger:97, a:Nalbach:10b} and numerically exact 
approaches based on the Feynman-Vernon influence functional.~\cite{a:Feynman:63}
Representations of the latter are the hierarchical equations of motion (HEOM) method~\cite{a:Tanimura:90,Tanimura:JCP:22} by Kubo and Tanimura and the  
quasi-adiabatic propagator path integral (QUAPI) method~\cite{a:Makri:95, a:Sim:97, a:Makri:95b, a:Makarov:94, a:Makri:95c} introduced by Makarov and Makri. 
While quantum master equation methods strongly rely on weak-coupling and Markovian approximations that can be reasonable for simple examples ~\cite{a:Moy:99, a:Holland:96},
most realistic systems of interest show long lived correlations and intermediate-to-strong coupling of the system and environment.
Under such conditions, numerically more involved methods like HEOM might be better suited, the latter being rather efficient for  relatively high temperatures and specific forms of spectral densities. It should be noted, that substantial recent development effort made more general forms of the environment accessible to the HEOM method.~\cite{Lambert:NComm:19,Tanimura:JCP:22}

On the other hand, the QUAPI method has proven itself as a robust and accurate method with broad general applicability. While the extension to multiple  independent
commuting baths is straightforward within the QUAPI method,~\cite{a:Nalbach:11b,a:Nalbach:10}
the situation changes for multiple  non-commuting baths.
An implementation of two independent baths, one general and one bath commuting with the system Hamiltonian~\cite{a:Nalbach:10b,a:Palm:18} has been presented  albeit requiring a  substantially higher numerical effort. 
While the original QUAPI scheme for a TLS coupled to single general bath scales as $N_\textrm{paths} = 2^{2N+2}$, with $N$ being the memory time cut-off, the extension of the QUAPI method to the case of one general and one commuting baths scales as $N_\textrm{paths} = 2^{4N+4}$, and the even more general two independent general baths case would scale as $N_\textrm{paths} = 2^{6N+6}$.~\cite{phd:Palm:20}


A range of approximate numerical techniques have been developed 
for the QUAPI method with a single general bath with the general aim to reduce the number of paths and associated memory requirements of the method. 
Example of such developments are filtering of the propagator functional,~\cite{a:Sim:01}, the small matrix decomposition of Feynman path amplitudes,~\cite{Makri:JCTC:21}, mask assisted coarse-graining  of the quasi adiabatic propagator path integral~\cite{a:Richter:17} and time-evolving matrix product operators,~\cite{a:Strathearn:18} that have made the approach successively numerically efficient for simulations of ever more complex systems and environments. 
The  main purpose of the current work is to transfer successful  methods to the two independent non-commuting bath implementation of QUAPI and to test 
their efficiency and applicability to mitigate its computational requirements. 
The remainder of the manuscript is structured as follows: 
In Section~\ref{sec:theory}, we 
recapture the model Hamiltonian and discuss the number and types of baths coupled to TLS. In Section~\ref{subsec:th_approx}, the numerical approximations 
of interested are presented. The results are shown and discussed in Section~\ref{sec:results}. We summarize our conclusions in Section~\ref{sec:concl}. 
The relevant information that has not been included into the main text is presented in an Appendix. All equations are given in atomic units.

\section{Theory\label{sec:theory}}

\subsection{Model Hamiltonian\label{subsec:th_model}}
In what follows, we adopt the scenario of two independent baths, one general and one commuting with the system Hamiltonian~\cite{a:Palm:18}
as reasonable compromise between the physical complexity it reflects and the associated numerical costs. 
We consider a generic system-bath Hamiltonian of the form:
{\small\begin{equation}
H_\textrm{tot} = H_\textrm{S} + H_\textrm{SB}^{x} + H_\textrm{SB}^{z} \,,
\label{eq:2baths_Htot}
\end{equation}}
where $H_\textrm{S} = \frac{\Delta}{2} \sigma_x$ is a symmetric TLS with the tunneling amplitude  $\Delta$ and $\sigma_x$ is the respective Pauli matrix.
$H_\textrm{SB}^{x}$ and $H_\textrm{SB}^{z}$ represent 
fluctuation sources, where
the first bath is interacting through the $\sigma_x$ Pauli matrix (the $x$-bath) and is chosen to commute with system Hamiltonian ($\left[H_S, \sigma_x \right] = 0$). Such a fluctuation source is known as a pure dephasing bath, 
since it leads to decoherence of the TLS eigenstates without causing population relaxation.
The second environment, coupled through the $\sigma_z$ Pauli matrix (the $z$-bath), is a general fluctuation source that cause both, decoherence and population relaxation. It is a non-commuting fluctuation source  with respect to the TLS Hamiltonian under investigation ($\left[H_S, \sigma_z \right] \neq 0$).
Both baths are assumed to be harmonic and independent of each other:
{\small\begin{equation}
\left\{
\begin{aligned}
&H_\textrm{SB}^{\mu} = \sum_j \frac{\left( P_j^\mu \right)^2}{2M_j^\mu} + \frac{M_j^\mu \left( \omega_j^\mu \right)^2}{2} \left( Q_j^\mu - \frac{\lambda_j^\mu}{M_j^\mu \left( \omega_j^\mu \right)^2} \sigma_{\mu} \right)^2 
\\
&\left[ Q_{j^\prime}^{\mu^\prime}, P_j^{\mu} \right] = i \, \delta_{j^\prime, j} \, \delta_{\mu^\prime, \mu}  \,,
\end{aligned}
\right.
\label{eq:SB}
\end{equation}}
here $\mu$ runs over bath indices $x$ and $z$, $Q_j^\mu$ and $P_j^\mu$ are position and conjugate momentum operators of $j$-th mode with frequency 
$\omega_j^\mu$ and coupled to the quantum system with coupling constant $\lambda_j^\mu$. 

At thermal equilibrium, all the relevant statistical information about a harmonic bath is incorporated in the spectral density:
{\small\begin{equation}
J_\mu (\omega) = \frac{1}{2} \sum_j \frac{\left( \lambda_j^\mu \right)^2}{M_j^\mu \omega_j^\mu} \, \delta(\omega-\omega_j^\mu) \to \gamma_\mu \, \omega \, \exp\left( -\frac{\omega}{\omega_c^\mu} \right) \,,
\label{eq:Jw}
\end{equation}}
that is considered to be of an Ohmic form with linear dependence at small frequencies and an exponential cut-off at the high frequency limit. For the purpose of the current work, 
we consider commuting and non-commuting baths with identical spectral densities, weakly coupled to the quantum system $\gamma_x = \gamma_z = \frac{1}{16}$ and 
relatively large cut-off frequencies $\omega_c^x = \omega_c^z = 10\Delta$ that result in  a quasi-Markovian limit $\omega_c^{x/z} >> \Delta$.

\subsection{QUAPI with a Single General Bath\label{subsec:th_gen_bath}}

A special case of the general Hamiltonian in Eq.~(\ref{eq:2baths_Htot}) is the 
case of a  single general bath, i.e., $\gamma_x = 0$. 
The matrix element of TLS reduced density matrix $\rho_S(t)$ is given in QUAPI theory~\cite{a:Makri:95, a:Sim:97, a:Makri:95b, a:Makarov:94, a:Makri:95c}: 
{\small\begin{multline}
\MXEL{\sigma_N^+}{\rho_S(t)}{\sigma_N^-} = \sum_{\sigma_{N-1}^\pm}\dots\sum_{\sigma_0^\pm} \MXEL{\sigma_0^+}{\rho_S(0)}{\sigma_0^-} \times
\\
\times G_S^z(\KET{\sigma_N^\pm}, \dots, \KET{\sigma_0^\pm}, N) \cdot I^z(\sigma_N^\pm, \dots, \sigma_0^\pm, N). 
\label{eq:1bath_gen}
\end{multline}}
Here, is discretized $t = N \cdot \Delta t$, $\KET{\sigma_N^\pm}$ and $\sigma_N^\pm$ are eigenvectors and eigenvalues of the system coupling 
operator $\sigma_z \KET{\sigma_N^\pm} = \sigma_N^\pm \KET{\sigma_N^\pm}$ (Eq.~\ref{eq:SB}) at time $N$. 
Forward and backward time propagation ``coordinates'' are denoted with ``$+$'' and ``$-$'', respectively.
{\small\begin{multline}
G_S^z(\KET{\sigma_N^\pm}, \dots, \KET{\sigma_0^\pm}, N) = 
\\
\prod_{j=1}^N \MXEL{\sigma_j^+}{e^{-i H_S \Delta t}}{\sigma_{j-1}^+} \MXEL{\sigma_{j-1}^-}{e^{i H_S \Delta t}}{\sigma_{j}^-}
\label{eq:GS_gen}
\end{multline}}
is a system propagator of the isolated system in the absence of coupling to any environment and the first term after the equality describes the system's forward propagation from time step $j-1$ to $j$ while the second term describes  backward propagation from $j$ to $j-1$

The bath affects the system dynamics through the Feynman-Vernon influence functional~\cite{a:Feynman:63}:
%
{\small\begin{multline}
I^z(\sigma_N^\pm, \dots, \sigma_0^\pm, N) = 
\\
\exp\left\{ -\sum_{j=0}^N \sum_{j^\prime = 0}^{j} \left( \sigma_j^+ -\sigma_j^- \right) \left( \eta_{jj^\prime}^z \sigma_{j^\prime}^+ - (\eta_{jj^\prime}^z)^\ast \sigma_{j^\prime}^- \right) \right\} \,,
\label{eq:Iz}
\end{multline}}
where influence coefficients~\cite{a:Sim:97} 
{\small\begin{equation}
\eta_{jj^\prime}^z = \int_{\left(j-\frac{1}{2}\right)\Delta t}^{\left(j+\frac{1}{2}\right)\Delta t} dt^\prime \int_{\left(j^\prime-\frac{1}{2}\right)\Delta t}^{\left(j^\prime+\frac{1}{2}\right)\Delta t} dt^{\prime\prime} \, C^z(t^\prime - t^{\prime\prime})
\end{equation}}
are the discrete analogue of the 
bath correlation function ${C^z(t^\prime - t^{\prime\prime}) = \langle \sum_k \lambda_k^z Q_k^z(t^\prime) \cdot \sum_{k^\prime} \lambda_{k^\prime}^z Q_{k^\prime}^z(t^{\prime\prime}) \rangle}$. 
These coefficients are responsible for non-Markovian effects by correlating any pair of time steps $j$ and $j^\prime$ in the past~\cite{a:Makri:95}. The explicit form 
for the influence coefficients for a single general bath is given in the Appendix~\ref{appdx:eta_gen}.

\subsection{QUAPI with a Single Pure Dephasing Bath\label{subsec:th_pure_dephas}}

Another limiting case of the general Hamiltonian in Eq.~(\ref{eq:2baths_Htot}) is given by the 
case of a single pure dephasing bath, i.e., $\gamma_z = 0$. 
The key difference to Section~\ref{subsec:th_gen_bath} is that the TLS Hamiltonian now commutes with coupling term (see above Sec.~\ref{subsec:th_model})
%
%
which renders the model essentially equivalent to the independent boson model~\cite{b:Breuer:02} that can be integrated out 
exactly. The general QUAPI 
numerical solution in Eq.~(\ref{eq:1bath_gen}) is still applicable with the only difference that there is no associated Trotter splitting error 
$O\left( \Delta t^2 \right)$. 
 The commutation relation $\left[H_S, \sigma_x \right] = 0$ 
implies that 
eigenvectors $\KET{\sigma_j}$ of the system coupling operator are also eigenvectors of system Hamiltonian $H_S \KET{\sigma_j} = \varepsilon_j \KET{\sigma_j}$ that leads to the simple eigenstate form 
of the isolated system propagator (cf. Eq.~\ref{eq:GS_gen}):
{\small\begin{equation}
G_S^x(\KET{\sigma_N^\pm}, \dots, \KET{\sigma_0^\pm}, N) =  e^{-i \left( \varepsilon_N^+ - \varepsilon_N^- \right) N\,\Delta t} \prod_{j=1}^N \delta_{\sigma_j^+, \sigma_0^+} \delta_{\sigma_j^-, \sigma_0^-} \,.
\label{eq:pd_Gs}
\end{equation}}
Due to the Kronecker deltas, the only contributing paths in the sum of Eq.~(\ref{eq:pd_Gs}) are the paths, where system ``coordinates'' 
$\sigma_j^\pm$ do not change over time. There are in total just $4$ such paths for a TLS. 
Eq.~(\ref{eq:pd_Gs}) may be solved analytically~\cite{a:Strathearn:17}:
{\small\begin{multline}
\MXEL{\sigma_N^+}{\rho_S(t)}{\sigma_N^-} = \MXEL{\sigma_N^+}{\rho_S(0)}{\sigma_N^-} \times 
\\
\times e^{-i \left( \varepsilon_N^+ -\varepsilon_N^- \right) t} \cdot e^{-\left( \sigma_N^+ -\sigma_N^- \right) \RE{L(t)}} \cdot e^{-i \left[ (\sigma_N^+)^2 - (\sigma_N^-)^2 \right] \IM{L(t)}} \,.
\label{eq:pd_rho}
\end{multline}}
The first exponent on the right side of Eq.~(\ref{eq:pd_rho}) describes coherent oscillation of non-diagonal reduced density matrix 
elements due to the system Hamiltonian. The second exponent 
is the exponential decay of the system coherences due to the interaction with bath. The last exponent reflects the renormalization effect of 
system coherences induced by the fluctuation source. Since the system coupling operator is a Pauli matrix, its eigenvalues are $\sigma_N^\pm = \{-1, +1 \}$ 
so that the frequency renormalization 
effect is absent. Function $L(t)$ is an integral analogue of the influence coefficients:
{\small\begin{equation}
L(t) = \sum_{j=0}^{N} \sum_{j^\prime=0}^j \eta_{jj^\prime} = \int_0^t dt^\prime \int_0^{t^\prime} dt^{\prime\prime} \, C(t^\prime - t^{\prime\prime}) \,.
\label{eq:int_eta}
\end{equation}}
In case of a pure dephasing bath, the detailed features of the bath correlation function are irrelevant for system dynamics. 
The corresponding dephasing rate depends on an integral intensity of the correlation function alone.

\subsection{QUAPI with Two Independent Non-commuting Baths\label{subsec:th_2baths}}

 The full Hamiltonian~(\ref{eq:2baths_Htot}) describes the general case of two independent environments that themselves are non-commuting ($\left[\sigma_x, \sigma_z \right] \neq 0$). A propagation scheme within QUAPI theory for the time evolution of the reduced density matrix was recently proposed:~\cite{a:Palm:18} 
{\small\begin{multline}
\MXEL{\sigma_{z,N}^+}{\rho_S(t)}{\sigma_{z,N}^-} = \sum_{\sigma_{x/z,N-1}^\pm}\dots\sum_{\sigma_{x/z,0}^\pm} \MXEL{\sigma_{z,0}^+}{\rho_S(0)}{\sigma_{z,0}^-} \times
\\
\times G_S^{xz}\left(\KET{\sigma_{x,N-1}^\pm}, \dots, \KET{\sigma_{x,0}^\pm}, \KET{\sigma_{z,N}^\pm}, \dots, \KET{\sigma_{z,0}^\pm}, N \right) \times
\\
\times I^x(\sigma_{x,N-1}^\pm, \dots, \sigma_{x,0}^\pm, N-1) \cdot I^z(\sigma_{z,N}^\pm, \dots, \sigma_{z,0}^\pm, N), 
\label{eq:2baths_rho}
\end{multline}}
where $\sigma_{x, N}^\pm$ and $\sigma_{z, N}^\pm$ denote the eigenvalues of the corresponding Pauli matrices for the forward (``$+$'') and backward (``$-$'') paths at time $t = N\cdot \Delta t$, and sums 
$\sum_{\sigma_{x/z, j}^\pm} = \sum_{\sigma_{x,j}^+} \sum_{\sigma_{x,j}^-} \sum_{\sigma_{z,j}^+} \sum_{\sigma_{z,j}^-}$ are written in the short notations. 

The influence functionals in Eq.~\ref{eq:2baths_rho} of the commuting $I^x$ and non-commuting $I^z$ baths appear factorized which is the consequence of the independent character of the environments. 
Both influence functionals take the form given in Eq.~(\ref{eq:Iz}). The corresponding influence 
coefficients $\eta_{jj^\prime}^z$ and $\eta_{jj^\prime}^x$, however, differ from each other, explicit expressions are given 
in Appendix~\ref{appdx:eta_gen} and 
\ref{appdx:eta_pd}, respectively. 

The isolated system propagator $G_S^{xz}$, in presence of two different sets of eigenvectors $\{ \KET{\sigma_x} \}$ and 
$\{ \KET{\sigma_z} \}$, takes the form: 
{\small\begin{equation}
\left\{
\begin{aligned}
&G_S^{xz}\left(\KET{\sigma_{x,N-1}^\pm}, \dots, \KET{\sigma_{x,0}^\pm}, \KET{\sigma_{z,N}^\pm}, \dots, \KET{\sigma_{z,0}^\pm}, N \right) = 
\\
&\qquad\qquad = \prod_{j=1}^N \tilde{G}_S^{xz} \left( \KET{\sigma_{z,j-1}^\pm} \to \KET{\sigma_{x,j-1}^\pm} \to \KET{\sigma_{z,j}^\pm} \right)
\\
&\tilde{G}_S^{xz} \left( \KET{\sigma_{z,j-1}^\pm} \to \KET{\sigma_{x,j-1}^\pm} \to \KET{\sigma_{z,j}^\pm} \right) = 
\\
&\qquad = \BRAKET{\sigma_{z,j}^+}{\sigma_{x,j-1}^+} e^{-i\varepsilon_{j-1}^+ \Delta t} \BRAKET{\sigma_{x,j-1}^+}{\sigma_{z,j-1}^+} \times
\\
&\qquad\qquad\qquad \times \BRAKET{\sigma_{z,j-1}^-}{\sigma_{x,j-1}^-} e^{i\varepsilon_{j-1}^- \Delta t} \BRAKET{\sigma_{x,j-1}^-}{\sigma_{z,j}^-}.
\end{aligned}
\right.
\label{eq:GS_2baths}
\end{equation}}
Note the difference between Eqs.~(\ref{eq:GS_gen}), (\ref{eq:pd_Gs}), and (\ref{eq:GS_2baths}). In case of a single general bath, 
the quantum system during a single time step 
$\Delta t$ is propagated between two eigenstates of system coupling operator at two subsequent times  
$\KET{\sigma_{j-1}^\pm} \to \KET{\sigma_{j}^\pm}$. For a single pure dephasing bath, Kronecker deltas lead to a direct propagation between initial and final 
states $\KET{\sigma_{0}^\pm} \to \KET{\sigma_{N}^\pm}$. In contrast, in the simultaneous presence of both baths, the system propagator during each time step $\Delta t$ jumps 
fort and back between the two sets of system coupling eigenvectors. Actual propagation between eigenvectors of a general bath $\sigma_z$ at subsequent time 
points is always mediated by transition to the eigenvectors of the pure dephasing bath $\KET{\sigma_{z,j-1}^\pm} \to \KET{\sigma_{x,j-1}^\pm} \to \KET{\sigma_{z,j}^\pm}$. 
This behavior eliminates the simple Markovian property observed for 
a single pure dephasing bath described in Section~\ref{subsec:th_pure_dephas}, and makes system dynamics more involved. The introduced coupling is expected to lead to deviations from the case of two weakly coupled independent environment.

\section{Numerical approximations\label{subsec:th_approx}}

The QUAPI method described by the expressions presented in Sections~\ref{subsec:th_gen_bath}-\ref{subsec:th_2baths} provides numerically exact solutions for the time evolution of the reduced density matrix with the only convergence parameter being the Trotter time step $\Delta t$. 
Nevertheless, a direct application of Eqs.~(\ref{eq:1bath_gen}) and (\ref{eq:2baths_rho}) to any physical system subject to non-Markovian effects is in practice limited by short propagation times $t=N\cdot \Delta t$ and very few quantum states $M$ due to the exponential increase of 
the number of paths, being $M^{2N+2}$ and $M^{2N_x+2N_z+4}$ for a single general bath and two baths, respectively. 
Numerical approximations have been introduced to reduce the associated memory and compute requirements of the QUAPI approach with a single general bath, by keeping the numerical accuracy of the path integral method.
In the following section we briefly describe such approximations and their advantages and possible 
limitations. We further discuss some implementation aspects with regard to a recent hash-facilitated implementation~\cite{a:Ovcharenko:25} of the MACGIC-QUAPI method.

\subsection{Sharp Memory Cut-off\label{subsec:mem_cut_off}}

The most basic approximation of the QUAPI scheme emerges from the observation that the bath correlation function $C(t^\prime-t^{\prime\prime})$ shows a decays to zero  
at large lag times~\cite{b:Weiss:12}. Thus, the distant time points are 
essentially uncorrelated which motivates the approximation of a finite memory time.~\cite{Makri:JMathPhys:95}
Here, the influence coefficients $\eta_{jj^\prime}$ are explicitly set to zero 
for times lags larger than a beforehand chosen bath memory time $t_\textrm{mem} = N_\textrm{mem} \cdot \Delta t$. 
If $t_\textrm{mem}$ is chosen large enough, so that all influence 
coefficients for longer lag times become negligible, the approximate QUAPI dynamics converges to the exact QUAPI limit.~\cite{a:Makri:95b, a:Makarov:94, a:Makri:95c},
At the same time, the introduction of a finite $t_\textrm{mem}$ 
substantially reduces the number of paths as the path considered during memory time are decoupled from the total propagation time ($M^{2N_\textrm{mem}+2} \ll M^{2N+2}$).
Although the sharp memory cut-off is generally not considered critical, it was recently pointed out for super-Ohmic spectral densities, that a cut-off in $t_\textrm{mem}$ results in qualitatively 
wrong asymptotic long time behavior~\cite{a:Vagov:11, a:Nahril:16}.

The single pure dephasing bath (cf. Sec.~\ref{subsec:th_pure_dephas}) allows to demonstrate the general problem of 
neglecting small long time correlations beyond the sharp memory cut-off.~\cite{a:Strathearn:17}
Let us fix memory time $t_\textrm{mem}$ 
so that $C(t > t_\textrm{mem}) = 0$ and exchange variables in Eq.~(\ref{eq:int_eta}):
{\small\begin{multline}
L(t) = \int_0^{t_\textrm{mem}} dt^\prime \int_0^{t^\prime} d\tau \, C(\tau) + \int_{t_\textrm{mem}}^{t} dt^\prime \int_0^{t_\textrm{mem}} d\tau \, C(\tau) = 
\\
= L(t_\textrm{mem}) + \dot{L}(t_\textrm{mem})\,(t-t_\textrm{mem}) \,.
\label{eq:LL}
\end{multline}}
Substitition Eq.~(\ref{eq:LL}) into Eq.~(\ref{eq:pd_rho}) gives:
{\small\begin{multline}
\MXEL{\sigma_N^+}{\rho_S(t)}{\sigma_N^-} = S \cdot e^{-i \left( \varepsilon_N^+ -\varepsilon_N^- \right) t} \times
\\
\times e^{-\left( \sigma_N^+ -\sigma_N^- \right) \RE{\dot{L}(t_\textrm{mem})} \cdot t} \cdot e^{-i \left[ (\sigma_N^+)^2 - (\sigma_N^-)^2 \right] \IM{\dot{L}(t_\textrm{mem})} \cdot t} \,,
\label{eq:pd_rho2}
\end{multline}}
where we combine all time independent factors into a single constant $S$, and dot denotes time derivative $\dot{L}(t) = \frac{d}{dt} L(t)$.

The two last exponents in Eq.~\ref{eq:pd_rho2} reflect the spurious behavior induced by the artificial restriction of memory time  $t_\textrm{mem}$. 
The last exponent introduces an erroneous of the system frequency. For coupling operators in the form of Pauli matrices this contribution is constant as $(\sigma_N^+)^2 = (\sigma_N^-)^2$. Errors induced by the second exponent are more severe. If $\dot{L}(t_\textrm{mem}) > 0$, the approximate dynamics will exhibit an exponential decay characterized by an overestimation of the decoherence rate. On the other hand, if $\dot{L}(t_\textrm{mem}) < 0$, system dynamics will  exponentially diverge. Although the latter behavior is mostly intrinsic to  
super-Ohmic environments with Ohmicity larger than some critical value (typically $s_\textrm{crit} = 2-3$)~\cite{a:Haikka:13}, in the case of multiple few correlated baths, 
the critical Ohmicity may decrease even to the sub-Ohmic range ($0 < s_\textrm{crit} \le 1$).~\cite{a:Strathearn:17}
Nevertheless, the erroneous decoherence rate is inevitable for any Ohmicity parameter which implies that 
typical convergence  checks with respect to to a finite cut-off of the memory time cut-off are merely assuring that $\dot{L}(t_\textrm{mem})$ becomes small enough so that spurious decoherence rate does not 
reveal itself on the timescale where system coherences are distinguishable from zero. 

Evidently, for system dynamics subject to a single general bath or two independent non-commuting baths, where the dynamics is not analytically solvable, the sharp cut-off of the memory time is of concern as well. Nevertheless, due to the involved numerical effort to extend memory time, the effects are harder to determine and investigate. 
One state-of-the-art approach 
consists in redefining the influence coefficients 
$\eta_{jj^\prime} \to \tilde \eta_{jj^\prime}$ for large correlation times $(j-j^\prime) \Delta t \approx t_\textrm{mem}$ in such a way that the integral function $L(t)$ in Eq.~(\ref{eq:int_eta}),  
itself obtained through the redefined coefficients, equals to its exact value.~\cite{a:Strathearn:17, a:Vagov:11, a:Nahril:16}
In this way integral properties of the 
redefined correlation function, and, as consequence, the long time dynamical behavior, are improved, albeit at the cost of wrong local correlations at large correlation times. Generally these long-time  
correlations are considered not to play an important role. 

While the effect of a sharply defined finite memory was successfully tested against numerical~\cite{a:Nalbach:11} and analytical~\cite{a:Thorwart:00} methods, the above discussion highlights that not all consequences of this 
approximation have been fully addressed. In the results section, we  will numerically demonstrate that the mask-assisted coarse graining 
method~\cite{a:Richter:17} (MACGIG-QUAPI) in its particular realization with hash maps (see Section~\ref{subsec:MACGIC}) provides a numerically accessible way to improve aforementioned integral properties and to investigate truncation effects of memory time without the need  to redefine the original influence coefficients.

\subsection{Hash-facilitated MACGIC-QUAPI\label{subsec:MACGIC}}

The bath correlation function $C(t^\prime - t^{\prime\prime})$ is typically characterized by a fast initial decay and a slower decaying tail for which a time uniform grid  with a time step $\Delta t$ is not  the most optimal sampling strategy. 
Denser temporal sampling for short correlation times and sparser sampling for longer times seems a better option. 
Such observation led to the proposal of a time non-uniform grid covering the bath memory time, called a mask.~\cite{a:Richter:17} 
The paths that coincide for the time points determined by the non-uniform mask but might differ on the  
the full uniform time grid spanning the bath memory time 
are considered as a single representative path.
In this way, the memory time is decoupled from the number of time points spanning the non-uniform mask and  
only a reduced set of representative paths is used for propagation which reduces the computational effort and memory requirements
($M^{2N_\textrm{mask}+2} << M^{2N_\textrm{mem}+2}$ and 
$M^{2N_\textrm{mask}^x+2N_\textrm{mask}^z+4} << M^{2N_\textrm{mem}^x+2N_\textrm{mem}^z+4}$ for a single general bath and  two independent non-commuting bath, respectively). 
The significant advantages of the MACGIC-QUAPI approach made biological 
complexes~\cite{a:Richter:17, a:Richter:19, a:Richter:20} accessible for QUAPI simulations.

In presence of the mask, the number of paths kept in computer memory is controlled by the size of the mask only and decoupled from memory time that can in principle be chosen equal to the total propagation time. This implies that even though only a reasonable small fraction of all possible paths is considered, each such path communicates back to initial time $t=0$  and it may be used for the exact calculation of the influence functional attributed to this particular path (see Eq.~(\ref{eq:Iz})).
Thus, we inherently account for the correct correlation function integral properties, as reflected by the function $L(t)$, and can eliminates the sharp memory cut-off problem at \textit{affordable} numerical costs. 

The mask-assisted coarse graining 
method~\cite{a:Richter:17} of MACGIC-QUAPI thus provides a significant advantage for long-time memory correlations that would be truncated by a sharp memory cut-off, an advantage that has not been systematically exploited so far. As we will demonstrate in Sec.~\ref{subsec:mem}, this allows us to strictly control the number of spawned paths by keeping the time integral limits in Eq.~(\ref{eq:int_eta}) equal to the full 
propagation time, thereby avoiding the uncertainties that are introduced due to the sharp cut-off of the memory time. 
In this way, the reliable numerical benchmark properties of the QUAPI  method are preserved. 

The numerical advantages of the MACGIC-QUAPI method come at the expense of comparing, reindexing, and finding equivalent paths that results in numerical overheads. A na\"ive look-up algorithm via a double-loop 
would scale quadratically  with the number of paths that makes it computationally unviable. The impoved look-up algorithm of Ref.~\citenum{a:Richter:17} was based on an additional pre-sorting step prior the actual merging step via the Radix-Sort algorithm  which improved the mask merging performance of the MACGIC-QUAPI algorithm. 
for a number of paths that fit into the computer memory of a single computing node. 
Overcoming this limitation, we have very recently developed a distributed memory implementation of the MACGIC-QUAPI algorithm 
based on a representation of paths as hash maps. The improved, hash-facilitated hMACGIC-QUAPI algorithm improves on the numerical scaling as 
$M^{2N+2} \to M^{2N}$ and $M^{2N^x+2N^y+4} \to M^{2N^x + 2N^y}$ for a single general bath and two independent non-commuting baths, respectively, and offers to distribute the memory requirements over multiple compute nodes via a hybrid distributed and shared memory parallelization. The interested reader is referred to Ref~\citenum{a:Ovcharenko:25} for implementation details and benchmark simulations for the single general bath case.

\subsection{Filtered Propagator Functional\label{subsec:FPF}}
An alternative strategy that allows in the redcution of paths used for propagation is given by the On-the-Fly Filtered Propagator Functional (OFPF) method~\cite{a:Sim:01}.
The method introduces a weight selection threshold $\theta$ such that any representative path, whose amplitude is smaller that this threshold 
$\vert \mathcal{A} \vert < \theta$, is eliminated from the calculation and does not participate in propagation. 
In the case of a single general bath, it was shown~\cite{a:Sim:97, a:Sim:96} 
that the fraction of high-weighted paths is in general small and neglecting paths with small amplitudes allows for a high accuracy in propagations while significantly reducing the number of paths. 
As filtering does not conserve the norm of the reduce density matrix, it is a standard practice to 
consider the norm of the reduced density matrix as convergence criterion. 

In Section~\ref{subsec:filtering} we explore the applicability of filtering for two independent non-commuting baths. Surprisingly, we find that (i) the dynamics starts diverging before the norm of the reduced density matrix norm starts to deviate substantially from $1$, and (ii) in contrast to the single bath scenario, for two baths there are no paths with low weights. Neglecting even a small fraction of paths with very small weight has drastic effects on the dynamics. 

\section{Results and discussion\label{sec:results}}

\subsection{Path Filtering\label{subsec:filtering}}

\begin{figure*}[ht!]
\includegraphics[height=0.23\textwidth]{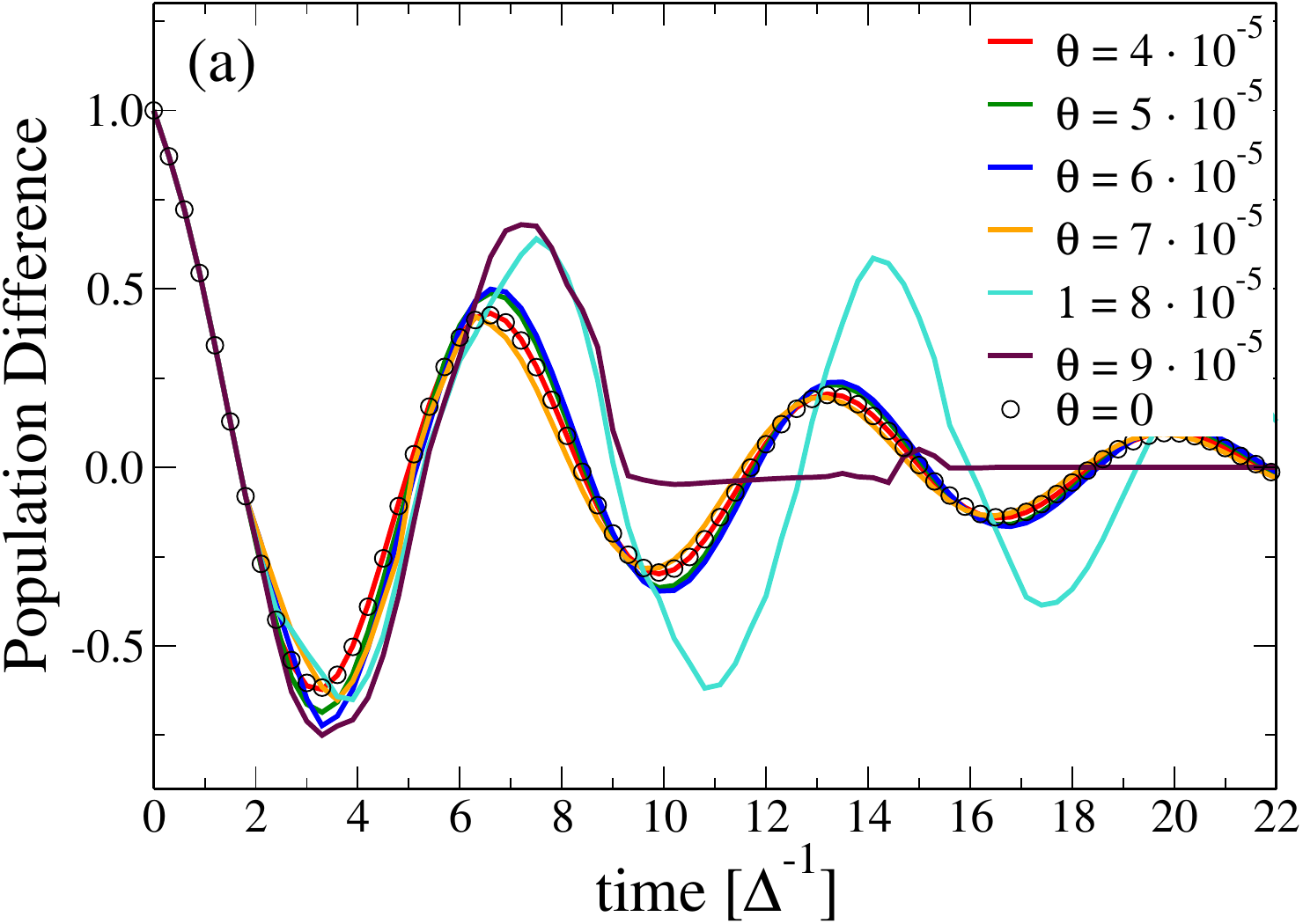}
\includegraphics[height=0.23\textwidth]{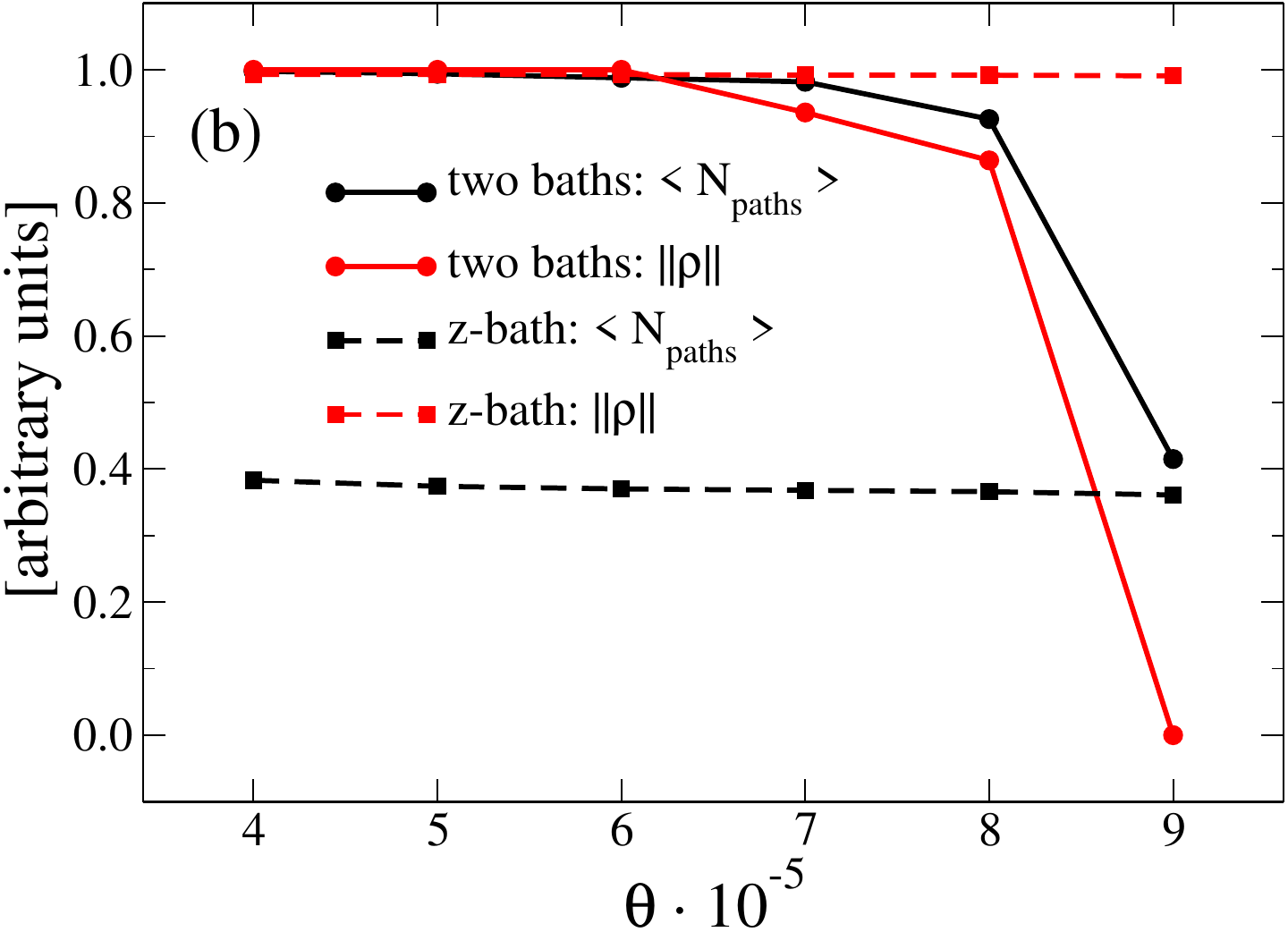}
\includegraphics[height=0.23\textwidth]{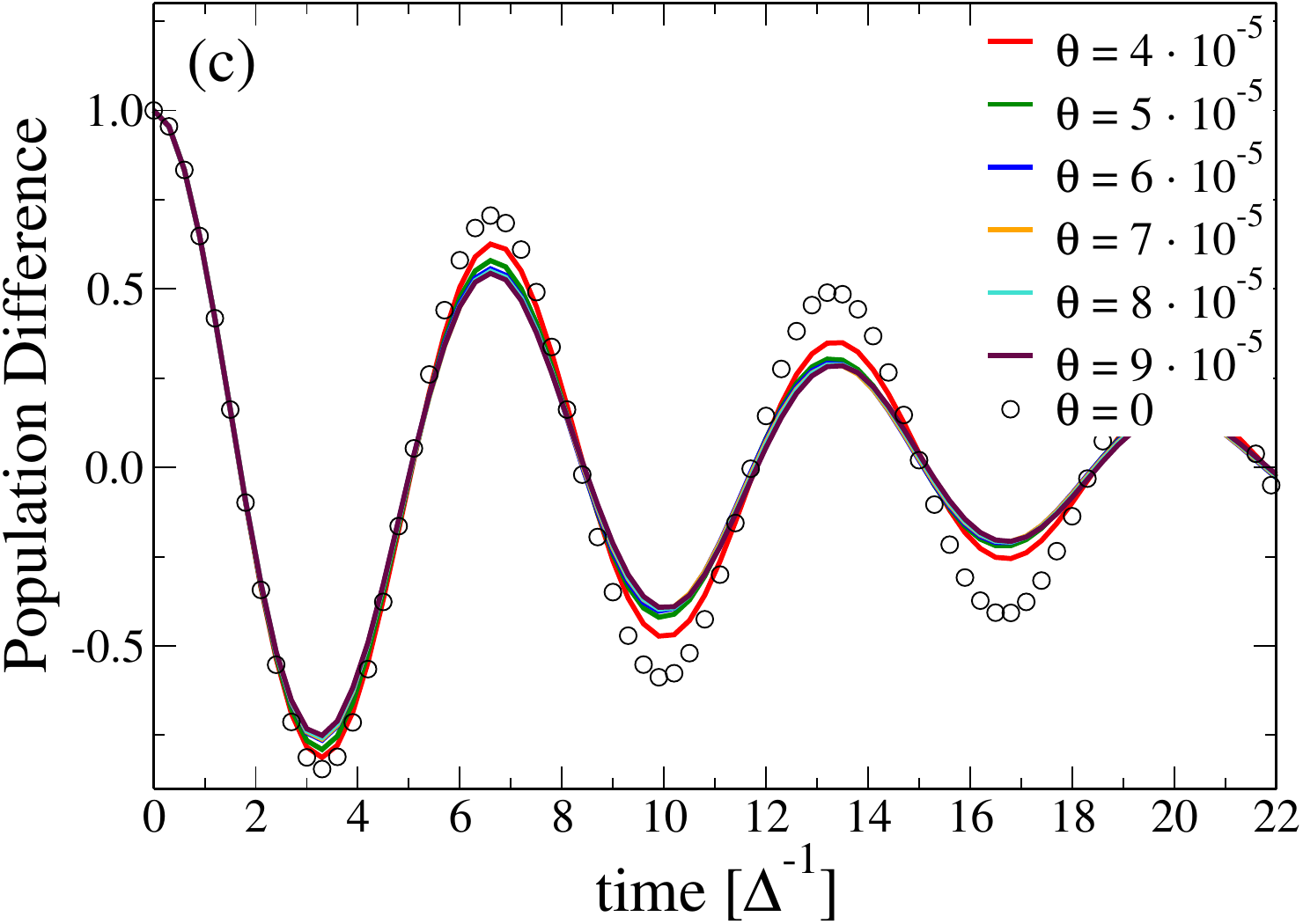}
\caption{Real-time dynamics for different filter threshold values $\theta$ is presented for the two independent non-commuting baths (a) and 
a single general bath (c). Panel~(b) shows fraction of paths used for propagation with given threshold value 
compared to the full QuAPI limit (black lines), the Norm of the reduced density matrix  
at the end of propagation $\parallel \rho(t = 35\Delta^{-1}) \parallel$ is presented with red lines.
\label{fig:2baths_filter}}
\end{figure*}
In the following we analyzed the applicability of the OFPF method for the dynamics of a TLS, weakly coupled to two independent non-commuting baths. Model parameters were taken from Ref.~\citenum{a:Palm:18} where the non-additive nature of both bath was revealed. We start by investigating of paths 
may be efficiently filtered out in presence of both fluctuating sources, simultaneously, offering a route for numerical savings.

Figure~\ref{fig:2baths_filter}~(a) demonstrates the real time dynamics of the quantum system calculated for 
different filter thresholds $\theta$, that are compared to the full QUAPI result without filtering (dotted data points, $\theta = 0$). 
Panel ~(b) gives the fraction of the paths during propagation compared to the number of path in the full QUAPI approach: 
$\langle N_\textrm{paths} \rangle = \frac{\int_0^{+\infty} N_\textrm{paths}^\theta (t) dt}{\int_0^{+\infty} N_\textrm{paths}^{\theta=0} (t) dt}$. Additionally, we depict the  
norm of the reduced density matrix at the end of the propagation time $\parallel \rho (t = 35\Delta^{-1}) \parallel$. 

For a threshold value $\theta = 4 \cdot 10^{-5}$ we find that the fully converged result can be obtained, 
the value leads to dynamics exactly on top of the full QUAPI benchmark.
At the same time, no paths are discarded, surprisingly, there are no paths with magnitude smaller than $4 \cdot 10^{-5}$. 
Lowering the threshold value to $\theta = 5 \cdot 10^{-5}$ and $\theta = 6 \cdot 10^{-5}$ results in 
noticeable deviations of the dynamics of the reduced density matrix from the reference while the number of paths participating in propagation 
is only slightly reduced compared to the full QUAPI limit. Thus, paths with small wheight  significantly contribute to the accuracy of the propagation and can hardly be neglected. 
Comparing the dynamics for $\theta = 5 \cdot 10^{-5}$ and $\theta = 6 \cdot 10^{-5}$, both are very close to each other. When converging the dynamics with respect to the filter threshold $\theta$, this can lead to the conclusion of convergence, even as deviations to the exact result persist.

A qualitative change takes place for threshold value  $\theta = 7 \cdot 10^{-5}$. 
Even though the dynamics appears reasonable accurate, compared to the converged case, the norm of the reduced density matrix starts do decrease substantially, reaching a value of $0.94$. Such decrease appears as even still  $98\%$ of all paths are employed for propagation.  
Thus, the norm of the reduced density matrix starts changing 
much earlier than a converged value of the filter threshold $\theta$ has been reached, demonstrating that $\parallel \rho \parallel$ does not serve as reliable convergence criterion in the tow bath case.
 Further increase of $\theta$ to $8\cdot 10^{-5}$ results in pathological 
dynamics and the norm of the density matrix drops to $0.86$ while still $93\%$ of the total number of paths are employed. 
Considerable savings in memory requirements are only possible from a threshold value of $9\cdot 10^{-5}$, which does not allows for accurate dynamics and in effect discards all the paths by the end of the simulation time. 
The results of Fig.~\ref{fig:2baths_filter} thus convincingly demonstrate that filtering based on the magnitude of contributing paths  is not applicable for the TLS interacting with two non-commuting baths, even at weak coupling and quasi-Markovian regimes. 

Figure~\ref{fig:2baths_filter}~(c) presents the respective dynamics of the TLS interacting with the single $z$-bath, the dashed lines in  Figure~\ref{fig:2baths_filter}~(b) present the ration of slected path for propagation and the norm of the reduced density matrix. 
We find that  magnitudes of paths are distributed more uniformly as compared to the case of two baths. 
In the range $4\cdot10^{-5}$ - $9\cdot10^{-5}$, less than $40\%$ of paths of path are employed for propagation which constitutes significant savings in computer memory. At the same time, the norm of the reduced density matrix remains close to $1$. 
The corresponding dynamics (Fig.~\ref{fig:2baths_filter}~(c)) shows is slowly varying with $\theta$ and  monotonous convergence for decreasing   $\theta$.
Such smooth dependence of the system dynamics on the number of paths 
allows to find the converged values of $\theta$ so that paths with small weights 
may be neglected without noticeable 
consequences for system dynamics. Such scenario does not apply for the two bath case.

Surprisingly, we have found that for the case with two non-commuting baths ($\theta = 4 \cdot 10^{-5}$), 
effectively no single path can be filtered out and filtering immediately starts to sacrifice the accuracy of the dynamics. Thus the paths with the smallest wheight (on the order of $4 \cdot 10^{-5}$) already significantly contribute. 
This implies that more refined techniques for reduction of the size of the propagator tensor, e.g. relying singular value decomposition~\cite{a:Strathearn:18, a:Orus:14}, could also face challenges and their applicability to two-bath cases appears is an open issue as well.


\subsection{Memory Cut-off\label{subsec:mem}}

\begin{figure*}
\includegraphics[width=0.49\textwidth]{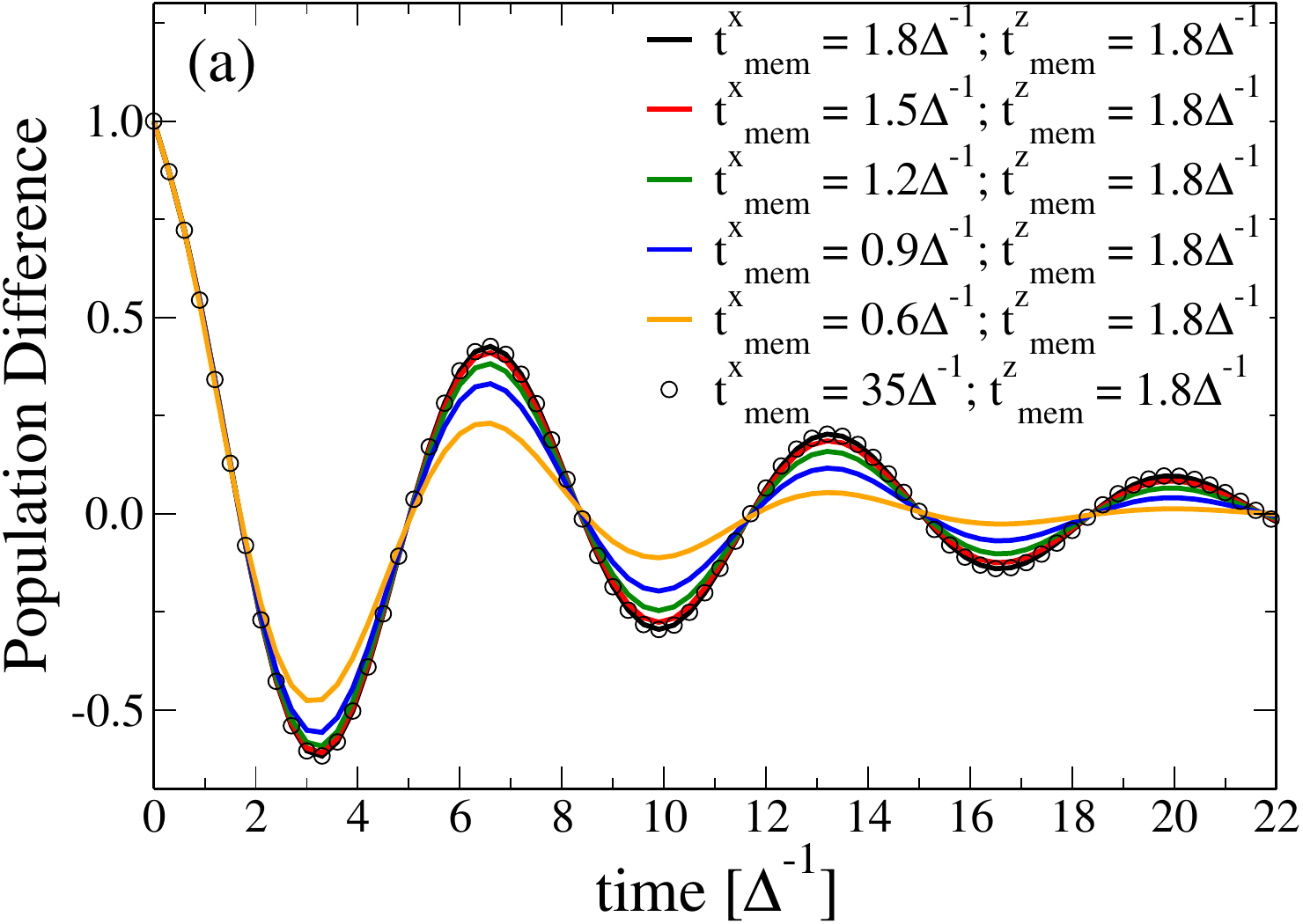}
\includegraphics[width=0.49\textwidth]{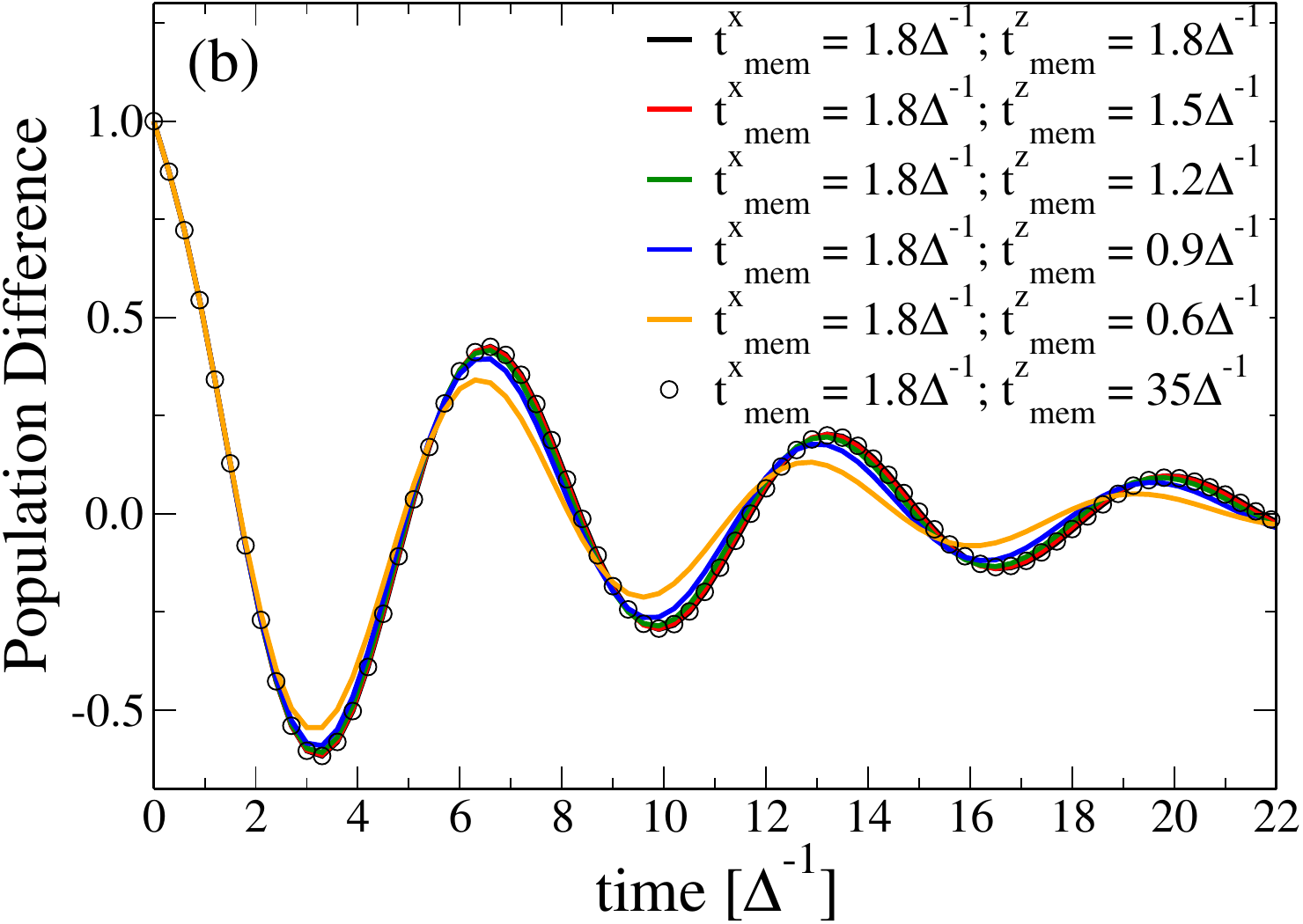}
\\
\includegraphics[width=0.49\textwidth]{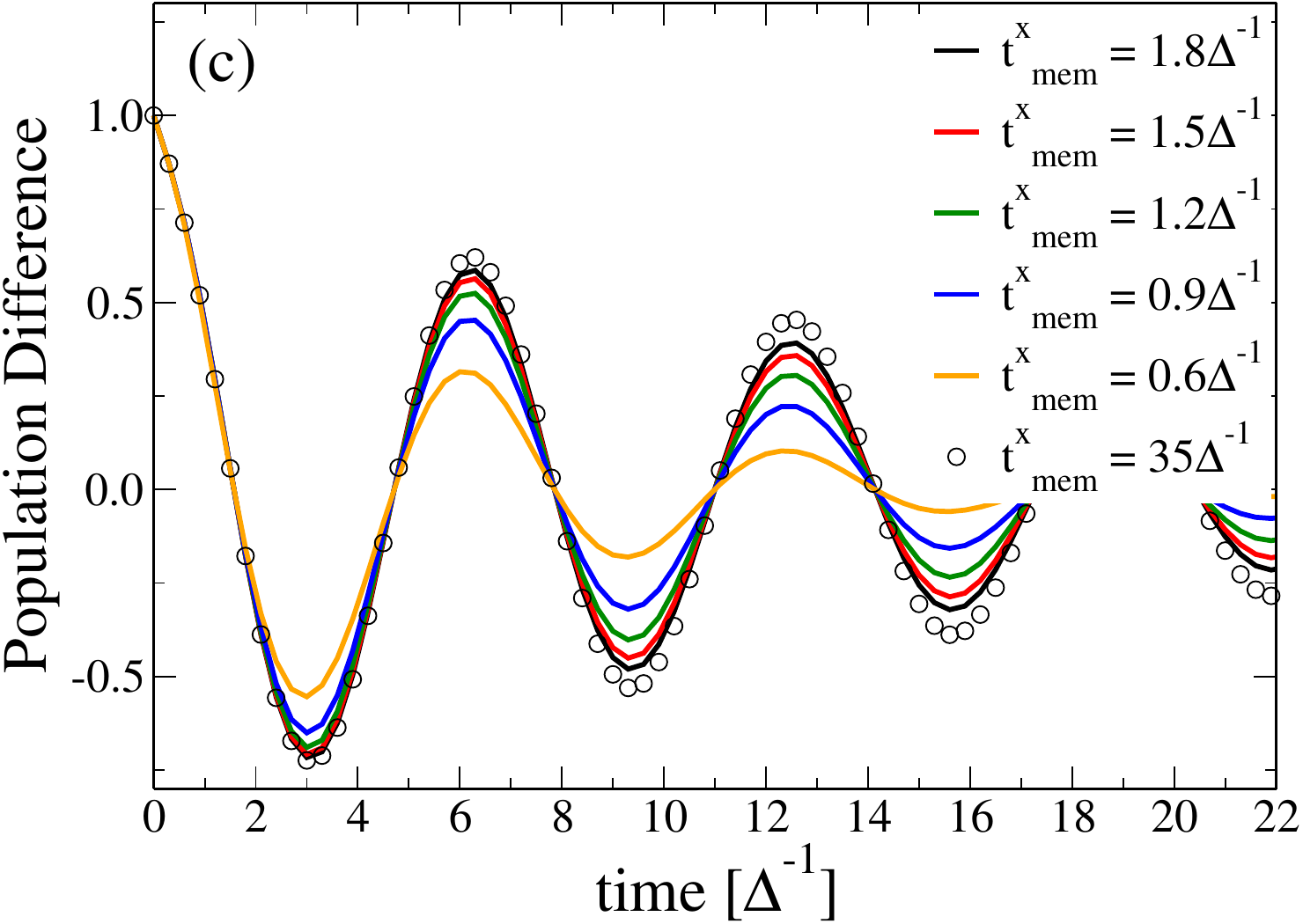}
\includegraphics[width=0.49\textwidth]{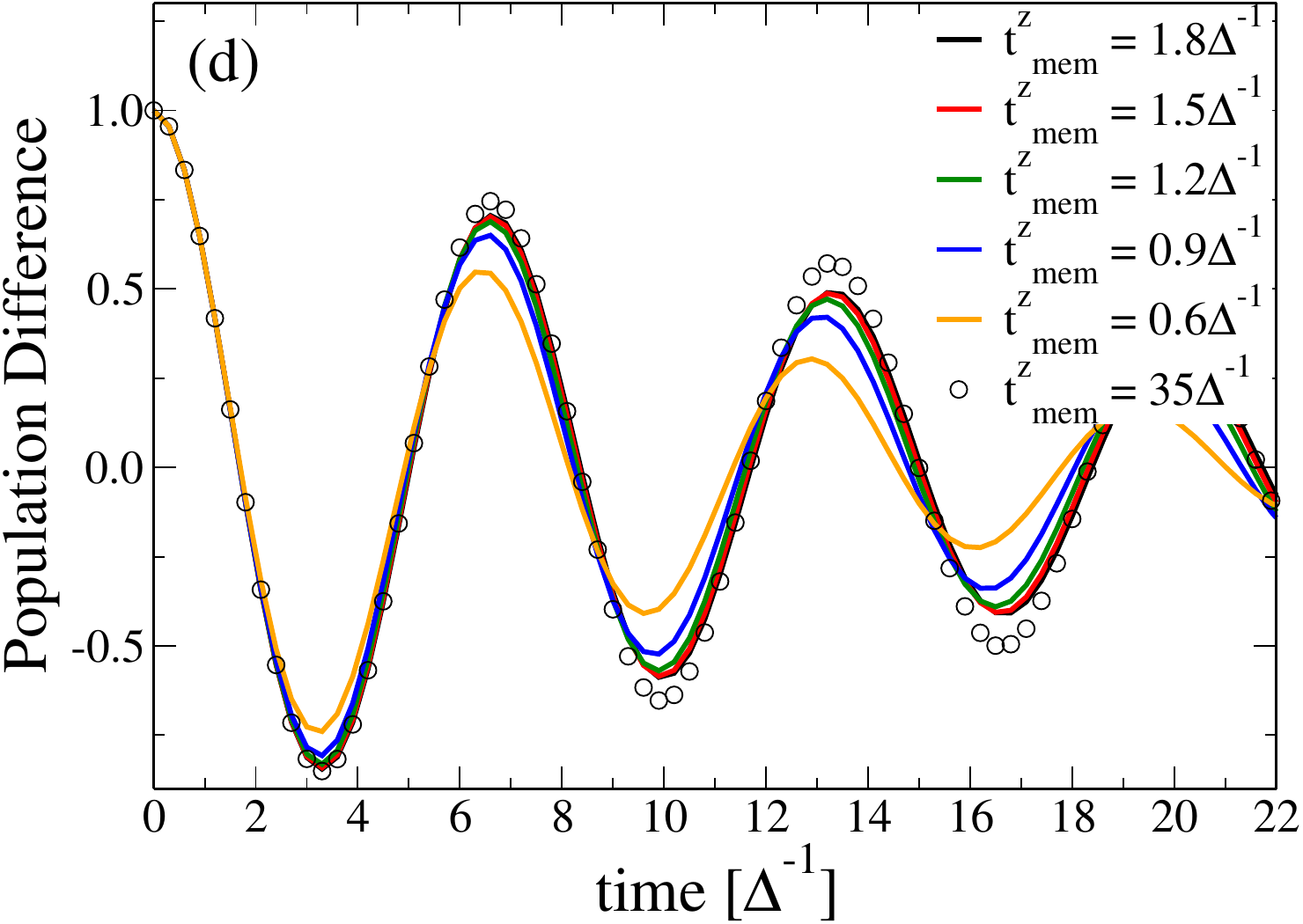}
\caption{Real-time dynamics of the TLS in presence of two independent, non-commuting  baths (a,b) and for single baths only (panels (c) and (d)). The mask in all simulations was chosen equal to the respective memory time, i.e., no mask was been employed.
Empty dots present numerically converged benchmark dynamics. For the benchmark calculations, a mask was employed that consists of the first $6$  time steps in memory time.  All simulation use the same time step 
$\Delta t = 0.3\Delta^{-1}$, no filtering was used.\label{fig:2baths_mem}
}
\end{figure*}

The overwhelming majority of performed QUAPI simulations relies on a cut-off of the memory time at a value for which long range correlations are presumed to be negligible.
As discussed in Sec.~\ref{subsec:mem_cut_off}, 
in some cases this may affect the system dynamics or even may lead to diverging behavior.
In the following we  numerically test approximation of a sharp memory cut-off applied to the TLS interacting with two independent, non-commuting  baths. 
For the two bath case,  the cut-off of the memory time may be controlled independently for each bath ($t_\textrm{mem}^x$ and $t_\textrm{mem}^z$).
 
The dynamics of the TLS is presented in 
Fig.~\ref{fig:2baths_mem}~(a,~b), the memory time  $t_\textrm{mem}^x = t_\textrm{mem}^z = 1.8\Delta^{-1}$ represents the converged dynamics (convergence tests are preseneted in the Appendix, Sec.~\ref{appdx:2baths}). 
First, we monotonically  decrease the memory time of the commuting bath by steps of time step $\Delta t$
(Fig.~\ref{fig:2baths_mem}~(a)) and
keep the memory time of the non-commuting bath fixed at the converged value. 
Initially (red and green lines) we find only small deviations from the converged dynamics.
Starting from a reduced memory time $t_\textrm{mem}^x = 0.9\Delta^{-1}$ 
(blue line), the differences to the converged dynamics become more apparent. While the dynamics still shows very similar underdamped oscillating behavior with the same frequency, the rate of decoherence  is substantially overestimated.

In Figure~\ref{fig:2baths_mem}~(b), the memory time of the $x$-bath is kept constant and while the memory time of the   $z$-bath is decreased. While the rate of decoherence is now only slightly affected, the frequency of the underdamped oscillations is now substantially shifted to higher frequencies. Overall, the deviations to the converged dynamics are smaller compared to Fig.~\ref{fig:2baths_mem}~(a) and memory times 
$t_\textrm{mem}^z > 0.9\Delta^{-1}$ (blue line) can be considered as converged. 
Thus, comparing Fig.~\ref{fig:2baths_mem} (a) and (b) we observed an asymmetric sensitivity of the TLS dynamics dynamics to cut-off of memory time for the  commuting and non-commuting bath.
For comparison, Fig.~\ref{fig:2baths_mem}~(c, d) demonstrates the dependence of the TLS dynamics on the cut-off of the memory time for the respective singe bath cases ((c): $x$-bath, (d): $z$-bath). 
Compared to the two-bath case, the sensitivity of a cut-off of the memory time is already reflected  in the dynamics with a single bath.
Such  behavior can be understood by taking into account that two baths represent independent fluctuating sources reasonably weakly coupled to the TLS.
The overall effect of both 
baths may thus be considered as a cumulative effect of each bath, separately: If the quantum coherent dynamics was originally more sensitive to the pure dephasing fluctuations than to the general bath, this property is also preserved in the simultaneous presence of both environments.

The effect of a sharp memory cut-off for the pure dephasing bath consists in an increased rate of  decoherence (cf. real part of the exponential in  Eq.~(\ref{eq:pd_rho2}) and there is no related frequency renormalization due to symmetric eigenvalues of system coupling operators ($\sigma_{x,j}^\pm = \pm 1$, see above).
Although an analytical solution similar to Eq.~(\ref{eq:pd_rho2}) does not exist for 
the $z$-bath, the effect of a finite memory time 
may be still understood in terms of an \textit{additional} spurious decoherence rate and oscillation frequency renormalization. It is observed that the latter  
converges faster than the former and that system frequency renormalization persists regardless of
the symmetric properties of system coupling operator eigenvalues ($\sigma_{z,j}^\pm = \pm 1$). 
This demonstrates the more complex behavior of the general bath compared to that of the pure dephasing bath.

As both baths are characterized by the same spectral densities $J_x(\omega) = J_z(\omega)$, corresponding integral functions $\dot{L^x}(t_\textrm{mem}^x)$ and $\dot{L^z}(t_\textrm{mem}^z)$ are equal. Nevertheless, spurious decoherence due to the limiting memory time is different. 
Thus, the different sensitivity of TLS dynamics with respect to the $x$- and $z$-baths is a consequence of the different 
system coupling part part in system-bath Hamiltonian (Eq.~\ref{eq:SB}) and not the bath correlation function itself. 
Figure~\ref{fig:2baths_mem} demonstrates that the convergence of the bath memory time is not only determined by 
the bath correlation function alone but depends on the correlation decay of of the bath correlation function and properties of the system coupling operator to the bath. 
The commutation relations of these operators with the system Hamiltonian (commuting or non-commuting) is thus a criterion strongly affecting the reduced TLS dynamics.
%

Figure~\ref{fig:2baths_mem}~(a-d) additionally presents the dynamics considering an extremely long memory time (empty dots, $t_\textrm{mem}^{x/z} = 35\Delta^{-1}$) as numerically converged benchmark simulations. 
Such long memory times equal to the 
total propagation time, are facilitated by using the mask assisted approach adopted for the two bath case (cf.  Sec.~\ref{subsec:MACGIC}).
 To investigate the effect of truncation of the memory time, we choose all initial mask points uniformly distributed on the time grid, i.e., essentially equivalent to the 
calculation without mask 
and finite memory cut-off ($t_\textrm{mem}^{x/z} = 6\cdot \Delta t$).
Once such a mask was fixed, 
we extended the memory time to the full propagation time.

An interesting observation is that, while in the two bath case the memory time of $1.8\Delta^{-1}$ is fully converged in comparison to the benchmark dynamics, there are still noticeable  deviations in case of the dynamics with a single bath (cf. Fig.~\ref{fig:2baths_mem}~(a) and (b) vs. (c) and (d)). Two interplaying  factors contribute to this effect:
(i) errors due to finite memory time add up during the course of propagation so that they become noticeable only at later times and 
(ii) the effective decoherence 
rate of two baths is more pronounced than that for each bath separately, leading to a faster decay of the coherence dynamics.  

Considering the long memory time ($t_\textrm{mem}^{x/z} = 35\Delta^{-1}$) via the the mask assisted approach requires to account for $N_\textrm{paths} = 2^{2\cdot 6+2\cdot 6} = 16.777.216$ paths.  
On the other hand, the extended memory time introduces long-time correlations between any two time points during the propagation for the particular paths stored in computer memory.
As far as the mask includes the same time delays 
as a finite memory time does, the number of paths and 
path coordinates within time span $t_\textrm{mem}$ are exactly the same in both cases. 
The main difference is that, while for finite memory time we save paths with a time span $t_\textrm{mem}$ only, for the mask assisted calculation with the extended memory time we save the full 
paths from the current time 
point with delay $j-j^\prime=0$ back to the start of simulation with delay $j-j^\prime=N$ equal to the full propagation time. This allows us to account for correlations of each path, reflected either in the  
integral function $L(t)$ of 
Eq.~(\ref{eq:int_eta}) or the influence functional $I^{x/z}(\sigma_N^\pm, \dots , \sigma_0^\pm, N)$ of Eq.~(\ref{eq:Iz}), in 
a numerically exact manner inherent to QUAPI without memory truncation. 

The mask assisted approach to the QUAPI method allows to evade the problem of truncation of memory time, but the realization comes  at the expense of keeping the full path length in memory  
instead of just smaller sections determined by a short memory time.  
In the particular case for the dynamics presented in Fig.~\ref{fig:2baths_mem}~(a,~b), the increase in memory requirements required $28.531~\textrm{GB} / 2.28~\textrm{GB} \approx 12.5$ times more computer memory.
While such an increase might appear substantial, in terms of the QUAPI memory requirements for extending the memory time, such increase is not decisive. For example, if we consider (no mask approximation) an 
increase of the memory time of either the $x-$ or $z$-bath by two time steps, i.e., $N_\textrm{mem}^{x/z} \to N_\textrm{mem}^{x/z}+2$, the 
respective memory requirements would increase by a factor of $2^4 = 16$. 
In terms of performance, 
extending the memory time in hMACGIC-QUPI simulations required only $1.3$ times longer wall time (empty dots in Fig.~\ref{fig:2baths_mem}) than the simulations with finite memory cut-off ($t_\textrm{mem}^{x/z} = 6\cdot \Delta t$, black lines in Fig.~\ref{fig:2baths_mem}). 

\begin{figure*}[ht!]
\includegraphics[height=0.32\textwidth]{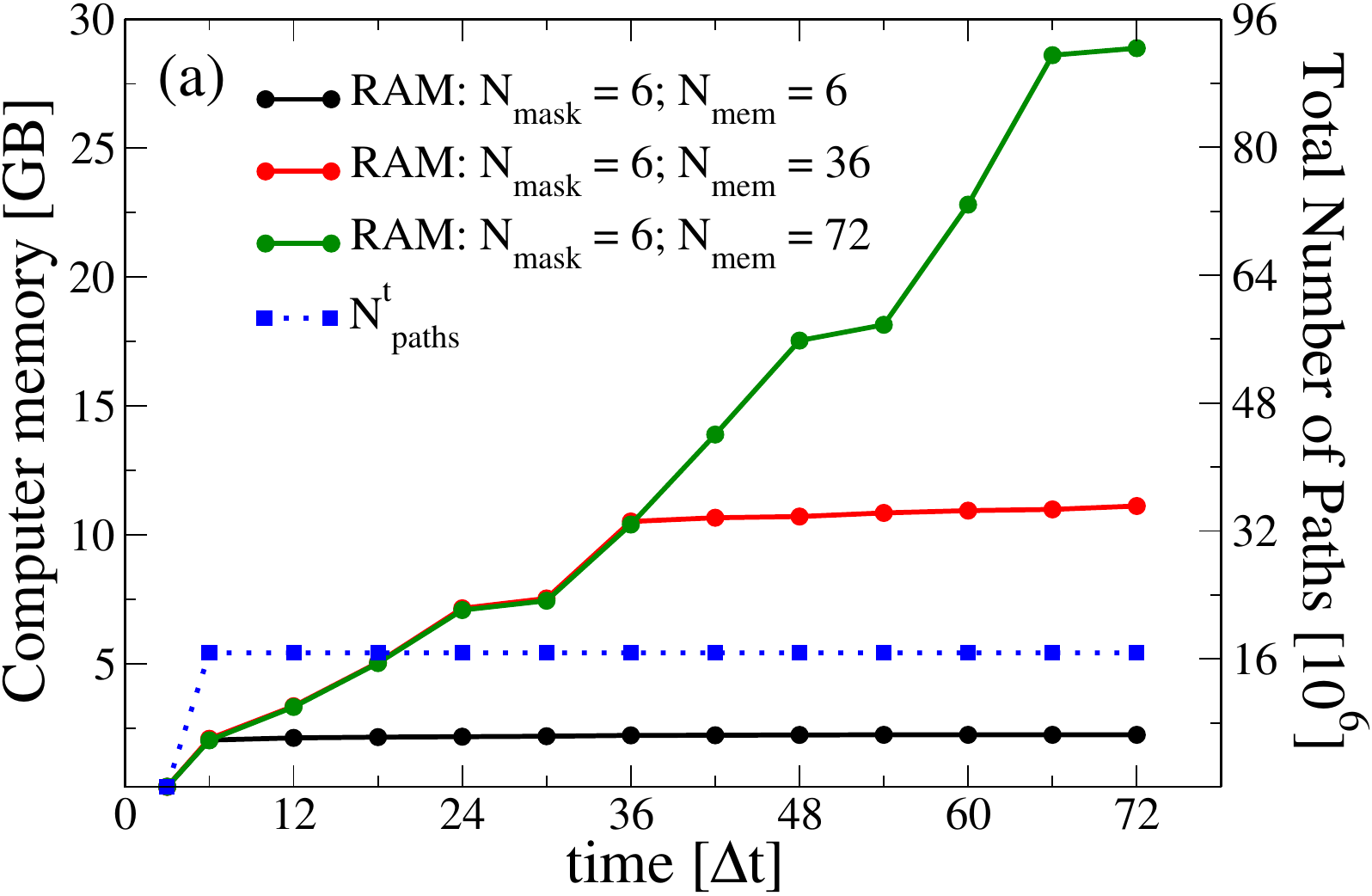}
\includegraphics[height=0.32\textwidth]{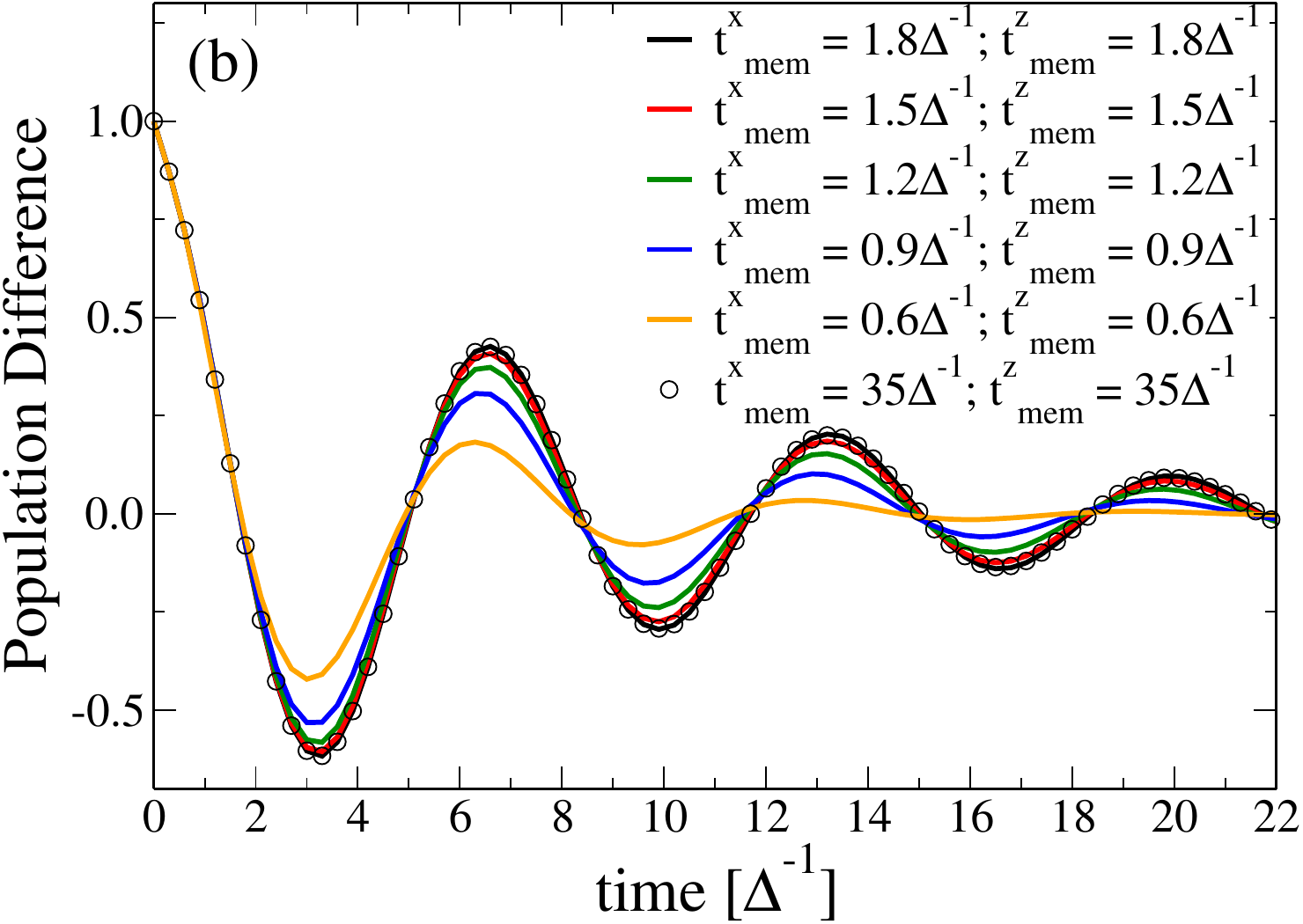}
\caption{
~(a) Average memory requirements (black, red, and green solid lines) and number of paths used for propagation (blue dotted line) 
in hMACGIC-QUAPI simulations in the presence of two independent non-commuting baths. Panel~(b) shows  
real-time dynamics in the presence of two independent non-commuting baths. Masks for both baths are chosen to be uniform and symmetric 
$N_\textrm{mask}^{x} = N_\textrm{mask}^{z} = N_\textrm{mask}$, memory times are also 
symmetric $N_\textrm{mem}^{x} = N_\textrm{mem}^{z} = N_\textrm{mem}$.  
Empty dots denotes benchmark dynamics where the mask consists of the $6$ 
most recent time steps. All simulations employed a time step $\Delta t = 0.3\Delta^{-1}$, no filtering was employed.
\label{fig:2baths_mem_both}}
\end{figure*}

The hMACGIC-QUAPI method allows to extend the bath memory time practically to arbitrary values. The chosen example of equating the memory and propagation time just serves as  most accurate, most memory consuming and performance demanding example.
The advantage of non-uniform mask consists in: 
(i) the decoupling of number of paths ($N_\textrm{mask}^{x/z}$) from 
the total correlation time of each path with itself in the past ($N_\textrm{mem}^{x/z}$) and (ii) the rigorous convergence checks with respect to  finite memory time in QUAPI simulations 
(fixing $N_\textrm{mask}^{x/z}$ and extending  
$t_\textrm{mem}^{x/z} \to t_\textrm{tot}$). 
Figure~\ref{fig:2baths_mem_both}~(a) presents the average memory requirement for the simulations with  
two independent, non-commuting baths  for different cut-off of the memory time. 
All simulations consider a fixed mask of size $N_\textrm{mask}^{x} = N_\textrm{mask}^{z} = 6$. The total number of paths considered in the simulations is given a blue line.
As expected, the number of paths grows rapidly until $t = N_\textrm{mask} \cdot \Delta t$,  after that value the total number of paths 
remains constant until the 
end of the propagation, irrespective of the memory cut-off. 

The same behavior is observed in simulations with a memory time equal to the mask size (Fig.~\ref{fig:2baths_mem_both}~(a), black line). 
The extension of the memory time to $N_\textrm{mem}^{x} = N_\textrm{mem}^{z} = 36$ time steps (red line) shows a different behaviors with regard to the memory requirements. Initially, until $t = N_\textrm{mask} \cdot \Delta t$, the number of paths and memory requirement grow, while in  the time range between $N_\textrm{mask} \cdot \Delta t$ and $N_\textrm{mem}\cdot \Delta t = 36\Delta t$, computer memory requirements  grow linearly due to the larger size of the paths to save. 
For the remainder of the propagation time $> N_\textrm{mem}\cdot \Delta t = 36\Delta t$, memory requirement stays approximately constant. 
If the memory time is extended to the full  propagation time (green line) 
memory grows approximate linear until the end of the propagation time.

In Figure~\ref{fig:2baths_mem_both}~(b) we investigate the effect of a symmetric reduction of the memory time for both baths.
The oscillation 
frequencies of the TLS dynamics becomes blue shifted for decreasing bath memory times, presumably due to the memory cut-off of the $z$-bath (cf. Fig.~\ref{fig:2baths_mem}~(b) and (d)). In addition, the rate of decoherence becomes overestimated resulting in a too fast damping of the coherent dynamics. 
The magnitude of the deviation  is larger than for the separate single bath cases, which is expected since both baths represent independent fluctuation sources and the overall effect on the TLS dynamics adds up. 

Comparing Figs.~\ref{fig:2baths_mem_both}~(b) and \ref{fig:2baths_mem}~(a,~b) suggests that 
a symmetric reduction of the memory time is 
not the most efficient strategy to optimize numerical effort while preserving accuracy of the dynamics.
While the simulations with $N_\textrm{paths} = 2^{2\cdot 5 + 2\cdot 5} = 1.048.576$ paths may be considered as converged (red line in Fig.~\ref{fig:2baths_mem_both}~(b)), Fig.~\ref{fig:2baths_mem}~(a,~b) suggests that a  asymmetric cut-off of the memory time ($5\Delta t;~3\Delta t$) is similarly accurate but more efficient 
 as $16$ times less paths ($N_\textrm{paths} = 2^{2\cdot 5 + 2\cdot 3} = 65.536$ paths) are considered. 
 We note that the asymmetry between memory times of both baths arises due to the different commutation relations of the system part of the system-bath coupling operators and system Hamiltonian. 
If different  spectral densities  are used for both baths, the asymmetry between memory times of baths can become even larger.

\subsection{Optimal Mask Choice}

\begin{figure*}
\includegraphics[width=0.49\textwidth]{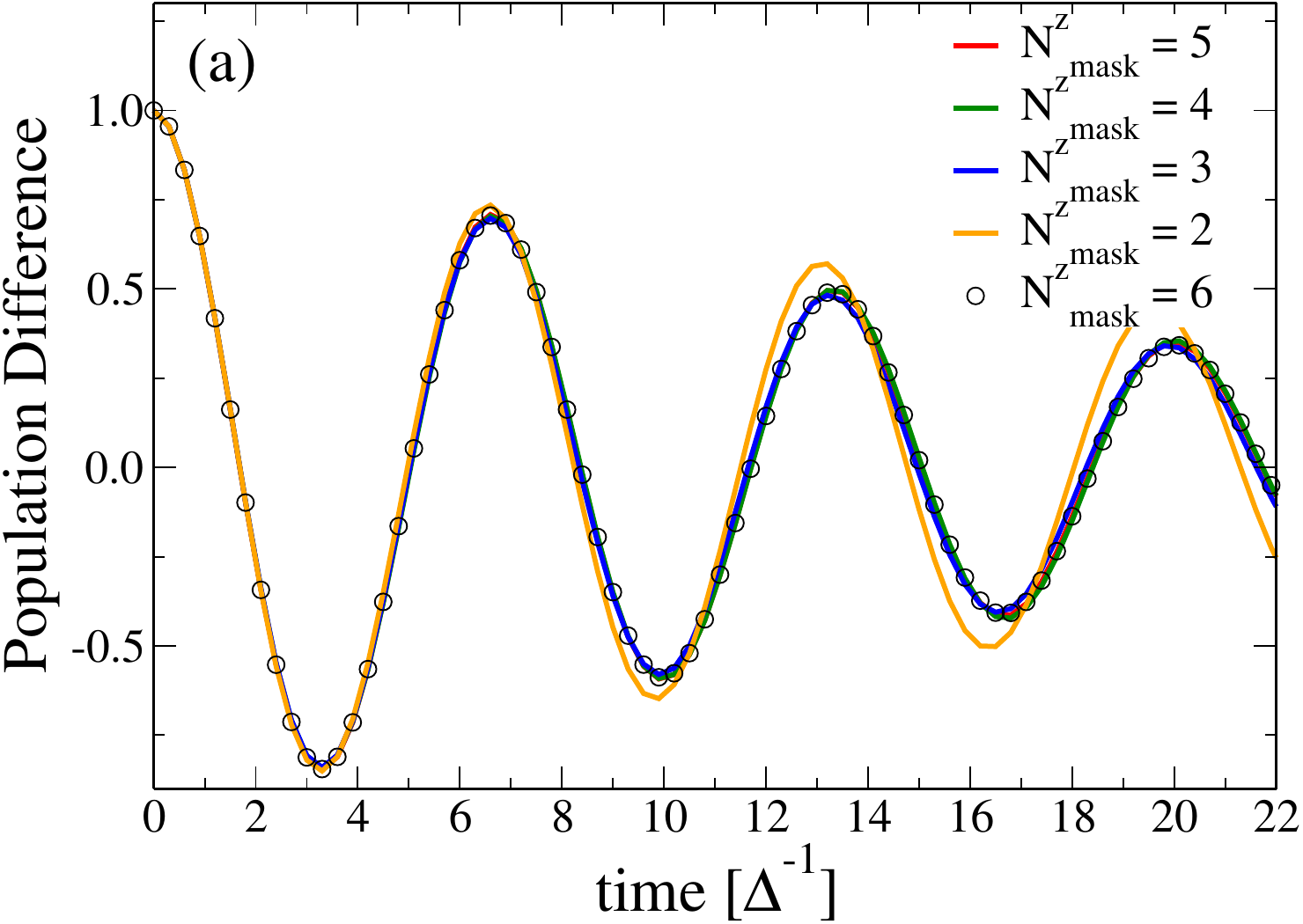}
\includegraphics[width=0.49\textwidth]{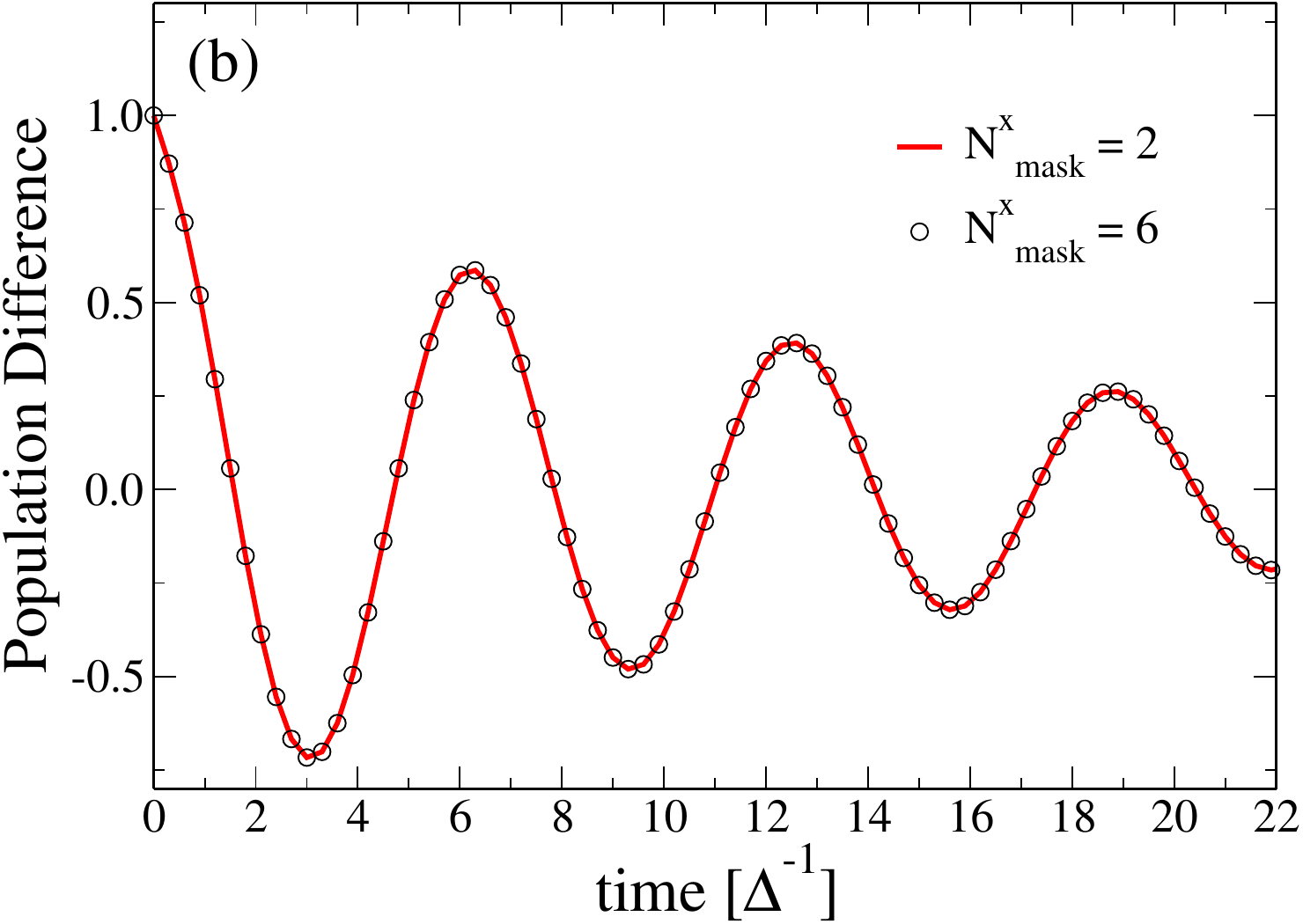}
\\
\includegraphics[width=0.49\textwidth]{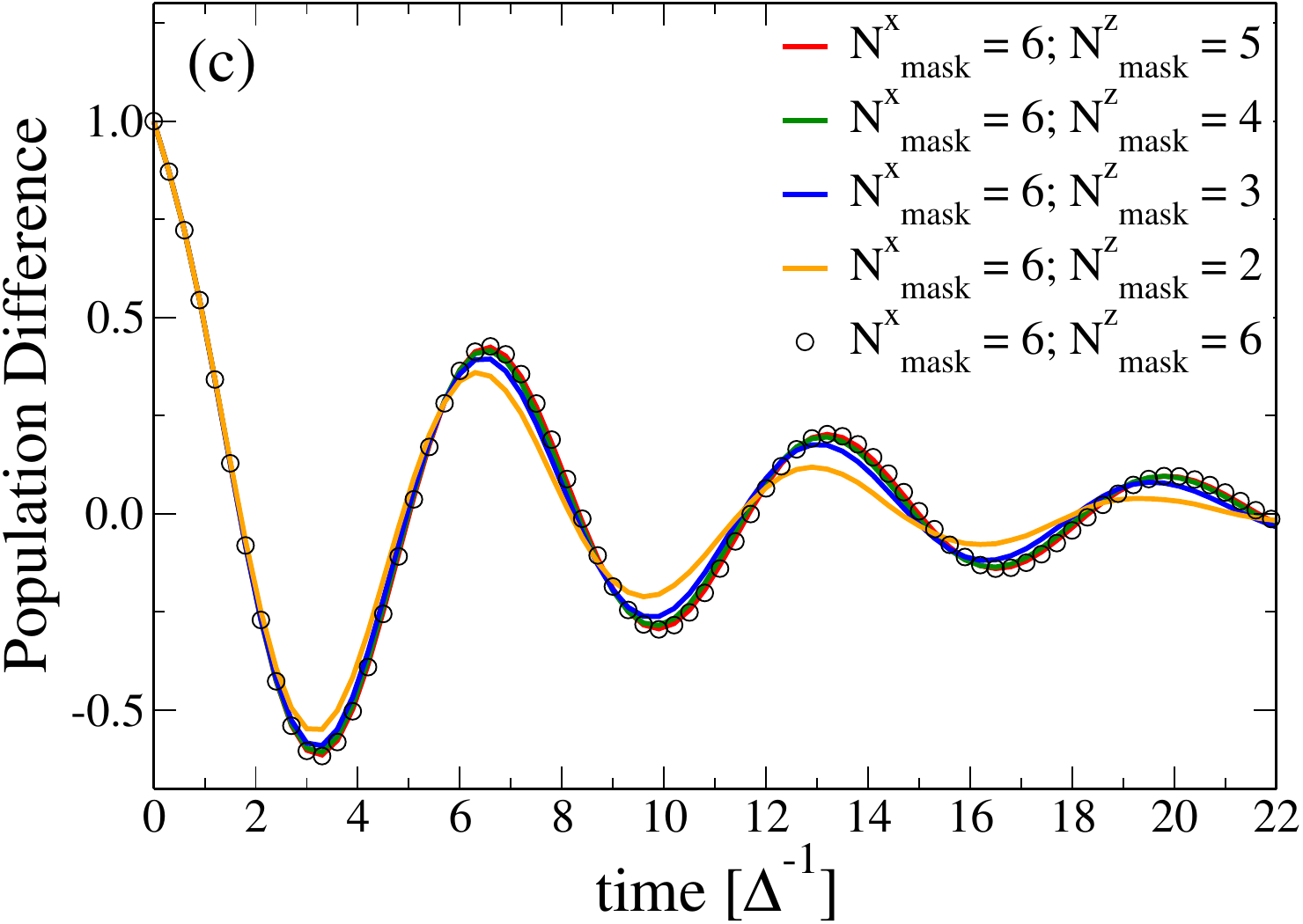}
\includegraphics[width=0.49\textwidth]{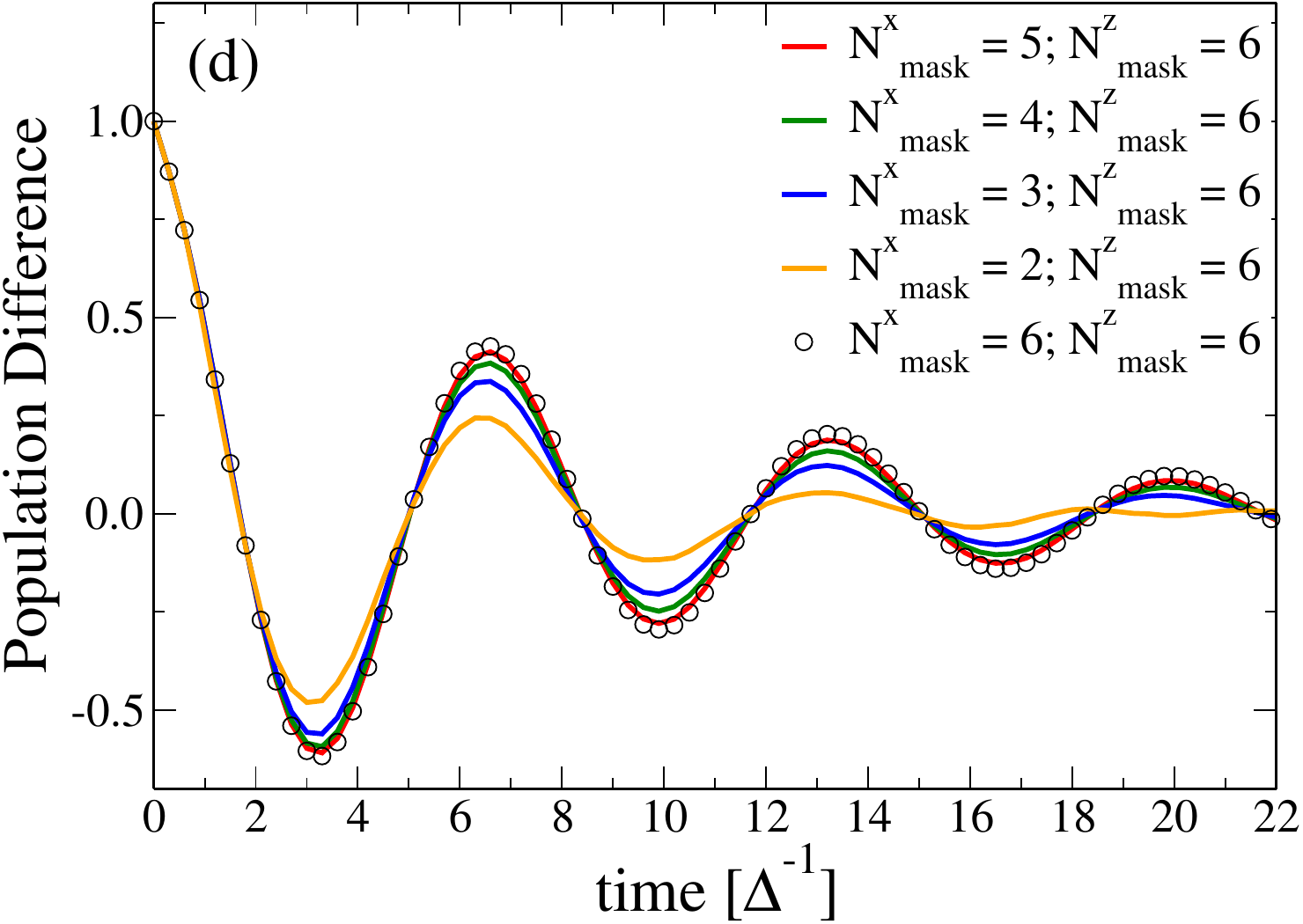}
\caption{Real-time dynamics of the TLS (a) for the singe general and (b) the single pure dephasing bath for varying mask sizes.
Panels (c) and (d) depict the dynamics of both baths simultaneously (pure depahsing and non-commuting). 
The memory time cut-off in all simulations is the converged value of $N_\textrm{mem}^x = N_\textrm{mem}^z = 6$ time steps.
For each $N_\textrm{mask}^{x/z}$, the optimal mask was obtained by finding the dynamics closest to the benchmark calculation with $N_\textrm{mask}^{x/z} = 6$ (empty dots). Time step $\Delta t = 0.3\Delta^{-1}$, no filtering was used.\label{fig:2baths_mask}}
\end{figure*}

So far we have employed a uniform mask
to decouple number of paths from the total time scan of the memory time. The original idea proposed in Ref.~\citenum{a:Richter:17} was a mask represented on a non-uniform time grid. 
In Fig.~\ref{fig:2baths_mask} we explore if this approach can be extended to two independent non-commuting baths.
First, we find that masks of different sizes can be exploited for 
a single general and a pure dephasing bath alone (Fig.~\ref{fig:2baths_mask} (a,b)). 
We therefore fixed the memory time for both baths  at $N_\textrm{mem}^{x/z} = 6$ time steps, i.e., a choice of $N_\textrm{mask}^{x/z} = 6$ corresponds to a full QUAPI simulation. 
We then calculated for all possible combinations 
$\binom{N_\textrm{mem}}{N_\textrm{mask}}$ the dynamics given mask of size $N_\textrm{mask}$
 to find the optimal mask yielding most accurate dynamics.
 For the dynamics in the presence of the single general bath (Fig.~\ref{fig:2baths_mask} (a)) it is easy to verify that convergence 
is reached at 
$N_\textrm{mask}^z = 3$  mask points 
which is smaller than for the uniform mask (cf. Fig.~\ref{fig:2baths_mem}~(d)). 
In the latter case, the converged memory time of 
$t_\textrm{mem}^z = 1.5\Delta^{-1}$ is reached for a mask  of size $N_\textrm{mask}^z = 5$. Thus, the non-uniform mask uses $4$ times less paths for convergence than the uniform one. 

Closer inspection of the optimal mask reveals that it contains the following time lag points $\mathcal{M}^z = \{ 0, 1, 3 \}$, 
where the value $0$ denotes the first  
correlation time $j-j^\prime$, i.e., intermediate delay $2$ is skipped. 
Such mask realization leads to a better coverage of the total memory time with 
a fewer number of mask points. 
The sampling follows common sense, with a denser sampling at early correlation times and sparser sampling at longer times, where the correlation function decays to zero.

Figure~\ref{fig:2baths_mask}~(b) demonstrate the TLS dynamics in presence of only the pure dephasing bath. In contrast to the general bath, a  
smaller mask of size $2$ is able to provide fully converged dynamics.
The optimal mask consists of the two most recent correlation time points $\mathcal{M}^x = \{ 0, 1 \}$, reflecting the short-time Markovian property of a pure dephasing bath
(cf. Sec.~\ref{subsec:th_pure_dephas}). Irrespective of the size of the mask, 
we always observe that exactly $4$ paths with non-zero amplitude contribute to the dynamics. Thus, not the number of paths but a converged memory time span that potentially can induce spurious effects in the dynamics (i.e. frequency renormalization and too fast rate of decoherence, see above Secs.~\ref{subsec:th_pure_dephas} and \ref{subsec:mem}) is decisive for convergence 
%

While the results presented in Fig.\ref{fig:2baths_mask}~(a,b) demonstrate that a non-uniform mask allows for substantial computational savings in the single bath case, 
it is an open question if such strategy can be directly transferred to the two bath case.
Considering, that in the current investigation both baths are independent and weakly coupled to the TLS, were both baths exert a near to cumulative effect (Sec.~\ref{subsec:mem}),
a viable strategy would be to construct the mask for the two bath base from the single bash constituents. If accurate, this would allow to add a pure dephasing bath almost for free because, in contrast to the general bath, it just needs to account for 4 path (two mask points) for convergence.

Figure~\ref{fig:2baths_mask}~(c) and (d) presents the dynamics of the TLS interactiong with  two non-commuting baths where in panel~(c) the mask size of the general bath is varied (keeping the mask of the pure dephasing bath unchanged) and in panel~(d) the mask size of the dephasing bath is varied. 
Concerning the mask size of the general bath, we observe that the TLS dynamics turns out to be slightly more sensitive to details of the mask  than for the single general bath cases (cf. panel (a)).
For converged dynamics at least $4$ mask points are required but the general convergence pattern of the mask for the $z$-bath are preserved
Inspection of the optimal 
mask yields $\mathcal{M}^z = \{ 0, 1, 2, 3 \}$, i.e., the best option is the uniform distribution of time delays. 
%

The situation changes for the pure dephasing bath (Fig.~\ref{fig:2baths_mask}~(d)). We observe that the dynamics is particularly sensitive to the size of the mask, for convergence at least a mask of size $N_\textrm{mask}^x = 5$ is required which is larger than that required for the general bath and contrary to the behavior for the pure dephasing bath alone  (cf. Fig.~\ref{fig:2baths_mask}~(b)).
Thus, the quasi-Markovian character of the pure dephasing bath is lost. 
Comparing the dynamics to the case of memory time cut-off (Fig.~\ref{fig:2baths_mem}~(a)) we observe similar differences which is a consequence of the optimal uniform mask neglecting same paths as for enlarging the memory time. 

The data presented in Fig.~\ref{fig:2baths_mask} (c,d) demonstrated that designing an optimal mask for computational efficiency requires special attention in the case of two independent non-commuting fluctuating environments as both baths behave in a non-additive way.
In particular, for two baths the independent character of baths is broken. Such complex interplay of both baths stems from the structure of the isolated system 
propagator (Eq.~(\ref{eq:GS_2baths})).
Propagation of the TLS for a single time step $\Delta t$ takes place between two eigenvectors of the system coupling operator of the general bath, i.e., 
$\KET{\sigma_{z, j-1}^\pm} - \rightarrow \KET{\sigma_{z, j}^\pm}$.
This propagation step involves as an intermediate the eigenvectors of the system coupling operator of the pure dephasing bath $\KET{\sigma_{x, j-1}^\pm}$. 
For a fixed set of initial and final eigenvectors, 
$\KET{\sigma_{z, j-1}^\pm} = \KET{\sigma_{z1}^\pm}$ and $\KET{\sigma_{z, j}^\pm} = \KET{\sigma_{z1}^\pm}$, there are four different paths corresponding to 
the different intermediate states: 
    \begin{enumerate}[label=(\roman*)]
        \item $\KET{\sigma_{z1}^\pm} \to \KET{\sigma_{x1}^+, \sigma_{x1}^-} \to \KET{\sigma_{z1}^\pm}$,
        \item $\KET{\sigma_{z1}^\pm} \to \KET{\sigma_{x1}^+, \sigma_{x2}^-} \to \KET{\sigma_{z1}^\pm}$,
        \item $\KET{\sigma_{z1}^\pm} \to \KET{\sigma_{x2}^+, \sigma_{x1}^-} \to \KET{\sigma_{z1}^\pm}$, and 
        \item $\KET{\sigma_{z1}^\pm} \to \KET{\sigma_{x2}^+, \sigma_{x2}^-} \to \KET{\sigma_{z1}^\pm}$.
    \end{enumerate}
The relative probabilities of these paths are related to 
the absolute value of the respective scalar products: 
{\small\begin{multline}
\left\vert G_S^{xz}\left( \KET{\sigma_{z1}^\pm} \to \KET{\sigma_{x1}^+, \sigma_{x2}^-} \to \KET{\sigma_{z1}^\pm} \right) \right\vert =
\\
= \left\vert \BRAKET{\sigma_{z1}^+}{\sigma_{x1}^+} \, \BRAKET{\sigma_{x1}^+}{\sigma_{z1}^+} \cdot \BRAKET{\sigma_{z1}^-}{\sigma_{x2}^-} \, \BRAKET{\sigma_{x2}^-}{\sigma_{z1}^-} \right\vert = \frac{1}{4}
\end{multline}}
and are independent on the intermediate states since $\vert \BRAKET{\sigma_{z, 1/2}}{\sigma_{x, 1/2}} \vert = \frac{1}{\sqrt{2}}$. 

If particular time points $j$ and $j-1$ 
belong to the mask $\mathcal{M}^z$ but the time point $j-1$ does not belong to the mask $\mathcal{M}^x$,
merging of these four paths by the mask involves the intermediate $\KET{\sigma_{x,j-1}^\pm}$ states to be combined into a single representative path determined by the largest amplitude. 
This, however, is an ill-defined operation because all four paths have exactly the same wheight. Such a merging operation would induce large errors during a single propagation time step  between the two consecutive states 
$\KET{\sigma_{z,j-1}^\pm} \to \KET{\sigma_{z,j}^\pm}$. 
According to this argumentation, the masks $\mathcal{M}^z$ and $\mathcal{M}^x$ of the two baths should be closely related to each other, concerning mask size and mask pattern
even though the baths represent qualitatively different fluctuation sources that might be represented by masks of very different sizes for each bath alone. 
\footnote{
One may argue that the isolated system propagator for time step $\Delta t$ should be independent on the number or type of the baths.
However, in QUAPI the structure of the system propagator is dictated by the number and  types of baths and the order in which they affect the dynamics of the quantum system
(cf. Eq.~(\ref{eq:GS_gen}), (\ref{eq:pd_Gs}), and (\ref{eq:GS_2baths}))
The structure of system propagator and the bath settings thus appear closely related, so that properties of the latter can, at least partially, affect the former. 
The finding of a \textit{balanced} requirement of masks as a consequence of particular structure of the system propagator follows from the analytic solution available for the pure dephasing path in $x$-direction alone where the  Kronecker delta in the system propagator  (Eq.~(\ref{eq:pd_Gs})) is a consequence thereof.
}

\begin{figure*}
\includegraphics[width=0.49\textwidth]{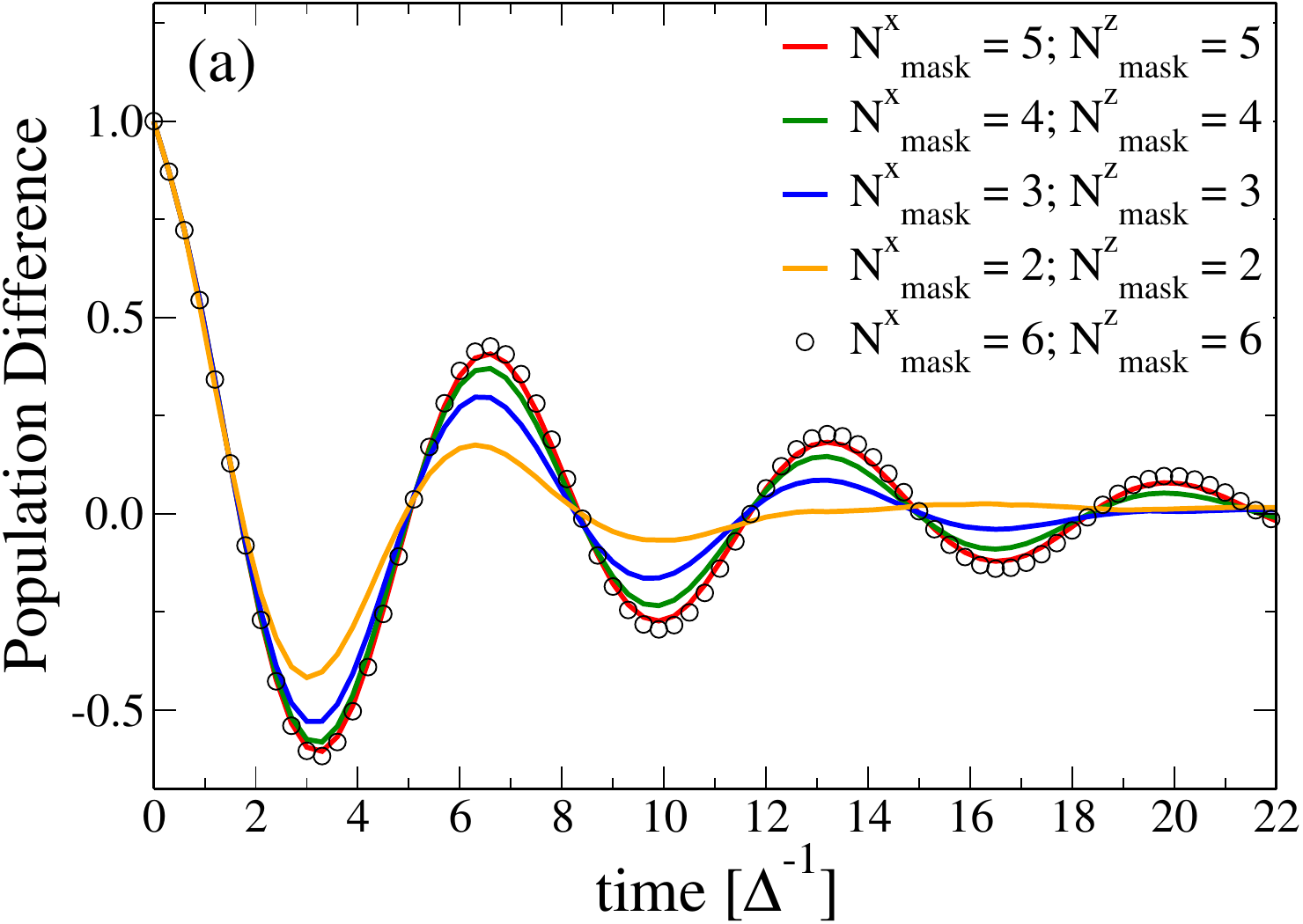}
\includegraphics[width=0.49\textwidth]{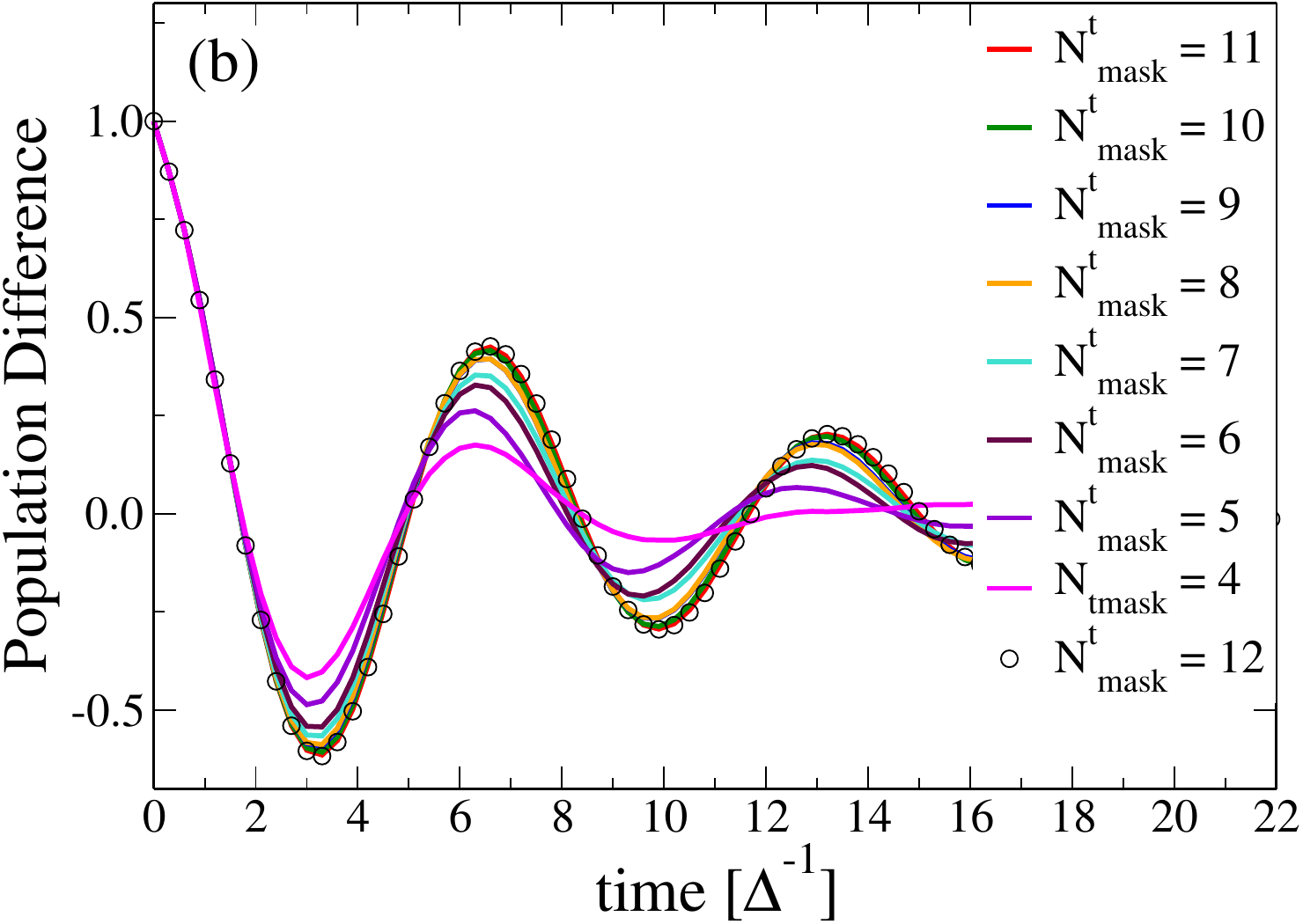}
\caption{Real-time dynamics of the TLS in presence of two independent non-commuting fluctuating environments for (a) symmetric variation of the mask size $N_\textrm{mask}^x + N_\textrm{mask}^z$ and (b) asymmetric varaiation with fixed total size of masks.   
Memory time in all simulations was the converged value of $N_\textrm{mem}^x = N_\textrm{mem}^z = 6$ time steps. 
For each mask size $N_\textrm{mask}^{x/z}$, all possible masks were evaluated and the optimal mask represents the dynamics closest to the benchmark simulation 
($N_\textrm{mask}^{x/z} = 6$, empty dots). 
$\Delta t = 0.3\Delta^{-1}$ and no filtering was used.\label{fig:2baths_mask_both}}
\end{figure*}

In Fig.~~\ref{fig:2baths_mask_both} we investigate the convergence behavior of the dynamics of the TLS with respect to (i) the sizes of symmetric masks $\mathcal{M}^z = \mathcal{M}^x$ (panel (a)), 
and (ii) for a total fixed mask size $N_\textrm{mask}^x + N_\textrm{mask}^z$ 
asymmetrically distributed between the masks for which the best realization was chosen.
$N_\textrm{mask}^x + N_\textrm{mask}^z$ (panel (b)). 
The comparison of the both data sets shows that symmetric masks are not the most optimal choice. Indeed, for the symmetric case, convergence is reached at ($5$; $5$) 
(panel (a), red line) requiring $10$ mask points. 
For the asymmetric case, convergence requires only $8$ mask points (panel (a), orange line).
Thus, superior masks are found for \textit{balanced} but not exactly equivalent realizations. 

Optimal mask combinations were derived by collecting all possible mask realizations for a given total number of mask points $N_\textrm{mask}^x + N_\textrm{mask}^z$ (see Table~\ref{tab:NxNz}).
\begin{table}
\caption{Different mask realizations for the real-time dynamics of the TLS shown in Figure~\ref{fig:2baths_mask_both}. Dark green color within each column indicates optimal 
mask realizations that 
result in closest dynamics compared to the benchmark simulation with $N_\textrm{mask}^x+N_\textrm{mask}^z = 12$. 
Light green color denotes mask realizations with reasonably close agreement, red colors indicate unsatisfactory dynamics. Reminder of combinations are indicated in black.
\label{tab:NxNz}}
\begin{ruledtabular}
\begin{tabular}{c|c|c|c|c|c|c|c}
$N_\textrm{mask}^x+N_\textrm{mask}^z$  & $5$ & $6$ & $7$ & $8$ & $9$ & $10$ & $11$ \\ 
\hline
& {\color{teal}($3$; $2$)} & {\color{teal}($4$; $2$)} & {\color{green}($5$; $2$)} & ($6$; $2$) & {\color{green}($6$; $3$)} & {\color{teal}($6$; $4$)} & {\color{teal}($6$; $5$)} \\
& {\color{green}($2$; $3$)} & {\color{green}($3$; $3$)} & {\color{teal}($4$; $3$)} & {\color{teal}($5$; $3$)} & {\color{teal}($5$; $4$)} & {\color{green}($5$; $5$)} & {\color{green}($5$; $6$)} \\
($N_\textrm{mask}^x$; $N_\textrm{mask}^z$) && {\color{red}($2$; $4$)} & {\color{green}($3$; $4$)} & {\color{green}($4$; $4$)} & {\color{green}($4$; $5$)} & {\color{green}($4$; $6$)} & \\
&&& {\color{red}($2$; $5$)} & ($3$; $5$) & {\color{red}($3$; $6$)} && \\
&&&& {\color{red}($2$; $6$)} &&& \\
\end{tabular}
\end{ruledtabular}
\end{table}
Masks were the classified according to four groups: (i) Optimal masks that resulted  closest dynamics compared to the benchmark simulation (dark green), 
(ii) close to optimal masks (light green), (iii) worst masks that features unsatisfactory dynamics compared to the benchmark, and 
(iv) the unspecified rest (black). 

Table~\ref{tab:NxNz} shows that for accurate dynamics, the representation of the $x$-bath is more important than representation of the $z$-bath.
Best realizations are generally characterized by $N_\textrm{mask}^x > N_\textrm{mask}^z$ (dark green color). In contrast,  unsatisfactory mask realizations 
are found for $N_\textrm{mask}^z$ being largest for fixed total number of points $N_\textrm{mask}^x + N_\textrm{mask}^z$ (see red color).
Mask realizations with good dynamics are found in vicinity of optimal ones (light green color) except for some highly unbalanced masks realizations, e.g. ($6$, $2$).
The described strategy allows to find the optimal mask that yields accurate dynamics  compared to the reference, offering substantial computational savings. Specifically, the number of paths considered for propagation can be substantially reduced  ($N_\textrm{paths} = 2^{2\cdot 6 + 2\cdot 6} = 16.777.216$ paths to $N_\textrm{paths} = 2^{2\cdot 5 + 2\cdot 3} = 65.536$ paths) without sacrificing accuracy substantially

The results of this section demonstrate that a general and a pure dephasing bath should not be treated on equivalent footing if both fluctuating sources act simultaneously. 
In contrast to the single bath case, were the description of pure dephasing is highly efficient, the accurate simultaneous description of a general and the pure dephasing environments requires a particular focus on the latter.  
Counter intuitive to the weak system-bath interaction strength, the influence of both baths appears to be strongly coupled and due to their non-additive nature, they can not be considered as independent environments anymore.

\section{Conclusions\label{sec:concl}}

A main challenge of highly accurate QUAPI simulations is the exponentially growing number of paths and associated requirements in computer memory for increasing bath correlation times.
This exponential wall is of even greater concern in presence of two independent non-commuting fluctuating environments, effectively making QUAPI simulations feasible only for small quantum systems and bath with very limited temporal correlations.
In the present work we investigate if approximations that increase the numerical efficiency of the QUAPI method with a single fluctuation source can be applied to two weakly-coupled independent and non-commuting baths, with one of them representing a pure dephasing environment. Such investigation paves the way to increasing the numerical efficiency of QUAPI simulations with two general bath  

We find that some commonly employed methods, like filtering of the paths according to their wheight, can not directly be transferred to the two bath case. It appears that the different paths are characterized by a high probability amplitude and neglecting even a small fraction of paths has drastic consequences for the dynamics of the reduced density matrix. 
Further, we explored how a finite truncation of the memory time affects the dynamics. We find that in the two bath case, the decoherence of the TLS is faster and that the effective 
memory is shortened compared to the situation where each bath acts independently alone. 
Moreover, for two baths with identical spectral densities an asymmetric cut-off of the memory time may be introduced  that was found to be more efficient than treating both bath with identical cut-off of the memory time.
Here we find that not only the correlation function but also system coupling operators in the system-bath operators are instrumental in determining the effective memory time of the individual baths.

We have further demonstrated that the mask assisted merging of paths that allows to decouple the numerical effort from memory time in the single bath case, can be applied in the case of two independent non-commuting baths as well, offering a route to computational savings.
Time non-uniform masks that have been found to be superior for the single bath case yield reasonable accuracy for the two bath case as well. Nevertheless, in the two bath case, uniform masks appear more efficient and allow to effectively decouple number of used paths during propagation (controlled by size of the mask) from 
the integral properties of the bath correlation function of each bath (given by the memory time cut-off). This approach offers increased flexibility to overcome limitations of a sharp memory cut-off in an exact manner inherent to QUAPI. 

Our work establishes the non-additive nature of two bath on the TSL dynamics. Even in the investigated weak coupling and quasi-Markovian regime, the independent character of the two baths may be misleading. While for some properties, such as the effective memory time, the intuitive logic of cumulative effects 
of the two weakly-coupled independent baths is confirmed, we have demonstrated that the quantum dynamics of the TLS appears particularly sensitive to the cut-off 
 of the memory time of a pure dephasing bath.
 This finding is contrary to findings for the general bath or in the presence of each bath alone. As a consequence, mask 
assisted path merging has to accurately capture the influence of the dephasing bath,  requiring a more accurate sampling of the 
$x$-bath via the mask than for the $z$-bath or in the respective single bath cases. 
Optimal representations of the mask thus appear with slight asymmetry towards the depahsing bath. The work demonstrates that the quasi-Markovian nature of a pure dephasing bath is lost, once another non-commuting source of fluctuations is present and that such non-additive effects require new strategies for numerical optimization.

\section*{Data Availability}
The data that support the findings of this study are available from the corresponding author upon reasonable request.

\begin{acknowledgments}
This research has received funding from the European Research Council under the European Unions Horizon 2020 and Horizon Europe Research and Innovation programs (Grant Agreements No. 802817 and  No. 101125590).
The authors would like to thank the Deutsche Forschungsgemeinschaft for ﬁnancial support [Cluster of Excellence \textit{e}-conversion (Grant No. EXC2089-390776260)].
The authors acknowledge the computational and data resources provided by the Leibniz Supercomputing Centre (www.lrz.de).
\end{acknowledgments}

\appendix

\section{Influence coefficients for a single general bath\label{appdx:eta_gen}}

The influence coefficients of the Feynman-Vernon influence functional are taken the from~\cite{a:Makri:95b}:
{\small\begin{equation}
I^z(\sigma_N^\pm, \dots, \sigma_0^\pm, N) = \exp\left\{ -\sum_{j=0}^N \sum_{j^\prime = 0}^{j} \left( \sigma_j^+ -\sigma_j^- \right) \left( \eta_{jj^\prime}^z \sigma_{j^\prime}^+ - (\eta_{jj^\prime}^z)^\ast \sigma_{j^\prime}^- \right) \right\} \,,
\end{equation}}
\begin{itemize}\itemindent=-5mm

\item[] If $0 < j^\prime < j < N$:
{\small\begin{multline}
\eta_{jj^\prime}^z = \int_{\left(j-\frac{1}{2}\right)\Delta t}^{\left(j+\frac{1}{2}\right)\Delta t} d\tau \int_{\left(j^\prime-\frac{1}{2}\right)\Delta t}^{\left(j^\prime+\frac{1}{2}\right)\Delta t} ds\ C^z(\tau - s) = 
\\
= \frac{2}{\pi} \int_{-\infty}^{+\infty} d\omega\ \frac{J_z(\omega)\, e^{ \frac{\beta\omega}{2} }}{ \omega^2 \sinh\left( \frac{\beta\omega}{2} \right) }\, \sin^2\left( \frac{\omega\Delta t}{2} \right)\, e^{ -i\omega\Delta t (j-j^\prime) }
\end{multline}}

\item[] If $0 < j^\prime = j < N$:
{\small\begin{multline}
\eta_{jj}^z = \int_{\left(j-\frac{1}{2}\right)\Delta t}^{\left(j+\frac{1}{2}\right)\Delta t} d\tau \int_{\left(j-\frac{1}{2}\right)\Delta t}^{\tau} ds\ C^z(\tau - s) = 
\\
\qquad = \frac{1}{2\pi} \int_{-\infty}^{+\infty} d\omega\ \frac{J_z(\omega)\, e^{ \frac{\beta\omega}{2} }}{ \omega^2 \sinh\left( \frac{\beta\omega}{2} \right) }\, \left( 1 - e^{ -i\omega\Delta t } \right) 
\end{multline}}

\item[] If $j^\prime = j = 0$:
{\small\begin{multline}
\eta_{00}^z = \int_{0}^{\frac{\Delta t}{2}} d\tau \int_{0}^{\tau} ds\ C^z(\tau - s) = 
\\
\qquad = \frac{1}{2\pi} \int_{-\infty}^{+\infty} d\omega\ \frac{J_z(\omega)\, e^{ \frac{\beta\omega}{2} }}{ \omega^2 \sinh\left( \frac{\beta\omega}{2} \right) }\, \left( 1 - e^{ -i\omega\frac{\Delta t}{2} } \right) 
\end{multline}}

\item[] If $j^\prime = j = N$:
{\small\begin{multline}
\eta_{NN}^z = \int_{\left(N-\frac{1}{2}\right)\Delta t}^{N \Delta t} d\tau \int_{\left(N-\frac{1}{2}\right)\Delta t}^{\tau} ds\ C^z(\tau - s) = 
\\
\qquad = \frac{1}{2\pi} \int_{-\infty}^{+\infty} d\omega\ \frac{J_z(\omega)\, e^{ \frac{\beta\omega}{2} }}{ \omega^2 \sinh\left( \frac{\beta\omega}{2} \right) }\, \left( 1 - e^{ -i\omega\frac{\Delta t}{2} } \right) 
\end{multline}}

\item[] If $0 = j^\prime < j < N$:
{\small\begin{multline}
\eta_{j0}^z = \int_{\left(j-\frac{1}{2}\right)\Delta t}^{\left(j+\frac{1}{2}\right)\Delta t} d\tau \int_{0}^{\frac{\Delta t}{2}} ds\ C^z(\tau - s) = 
\\
= \frac{2}{\pi} \int_{-\infty}^{+\infty} d\omega\ \frac{J_z(\omega)\, e^{ \frac{\beta\omega}{2} }}{ \omega^2 \sinh\left( \frac{\beta\omega}{2} \right) }\, \sin\left( \frac{\omega\Delta t}{4} \right)\, \sin\left( \frac{\omega\Delta t}{2} \right)\, e^{ -i\omega\Delta t \left( j-\frac{1}{4} \right) }
\end{multline}}

\item[] If $0 < j^\prime < j = N$:
{\small\begin{multline}
\eta_{Nj^\prime}^z = \int_{\left(N-\frac{1}{2}\right)\Delta t}^{N \Delta t} d\tau \int_{\left(j^\prime-\frac{1}{2}\right)\Delta t }^{ \left(j^\prime+\frac{1}{2}\right)\Delta t } ds\ C^z(\tau - s) = 
\\
= \frac{2}{\pi} \int_{-\infty}^{+\infty} d\omega\ \frac{J_z(\omega)\, e^{ \frac{\beta\omega}{2} }}{ \omega^2 \sinh\left( \frac{\beta\omega}{2} \right) }\, \sin\left( \frac{\omega\Delta t}{4} \right)\, \sin\left( \frac{\omega\Delta t}{2} \right)\, e^{ -i\omega\Delta t \left( N - j^\prime -\frac{1}{4} \right) }
\end{multline}}

\item[] If $0 = j^\prime < j = N$:
{\small\begin{multline}
\eta_{N0}^z = \int_{\left(N-\frac{1}{2}\right)\Delta t}^{N \Delta t} d\tau \int_{0}^{\frac{\Delta t}{2}} ds\ C^z(\tau - s) = 
\\
= \frac{2}{\pi} \int_{-\infty}^{+\infty} d\omega\ \frac{J_z(\omega)\, e^{ \frac{\beta\omega}{2} }}{ \omega^2 \sinh\left( \frac{\beta\omega}{2} \right) }\, \sin^2\left( \frac{\omega\Delta t}{4} \right)\, e^{ -i\omega\Delta t \left( N-\frac{1}{2} \right) }
\end{multline}}

\end{itemize}

\section{Influence coefficients for two independent non-commuting baths with one of the bath being a pure dephasing bath\label{appdx:eta_pd}}

The influence functional and influence coefficients for the non-commuting $z$-bath are exactly the same as that for the single general 
$z$-bath presented in Appendix~\ref{appdx:eta_gen}.

Influence coefficients for the commuting $x$-bath are taken from~\cite{a:Palm:18}:
{\small\begin{equation}
I^x(\sigma_N^\pm, \dots, \sigma_0^\pm, N-1) = \exp\left\{ -\sum_{j=0}^{N-1} \sum_{j^\prime = 0}^{j} \left( \sigma_j^+ -\sigma_j^- \right) \left( \eta_{jj^\prime}^x \sigma_{j^\prime}^+ - (\eta_{jj^\prime}^x)^\ast \sigma_{j^\prime}^- \right) \right\} 
\end{equation}}
\begin{itemize}\itemindent=-5mm

\item[] If $j^\prime < j$:
{\small\begin{multline}
\eta_{jj^\prime}^x = \int_{j \Delta t}^{\left(j+1\right)\Delta t} d\tau \int_{j^\prime \Delta t}^{\left(j^\prime+1\right)\Delta t} ds\ C^x(\tau - s) = 
\\
= \frac{2}{\pi} \int_{-\infty}^{+\infty} d\omega\ \frac{J_x(\omega)\, e^{ \frac{\beta\omega}{2} }}{ \omega^2 \sinh\left( \frac{\beta\omega}{2} \right) }\, \sin^2\left( \frac{\omega\Delta t}{2} \right)\, e^{ -i\omega\Delta t (j-j^\prime) }
\end{multline}}

\item[] If $j^\prime = j$:
{\small\begin{multline}
\eta_{jj}^x = \int_{j \Delta t}^{\left(j+1\right)\Delta t} d\tau \int_{j \Delta t}^{\tau} ds\ C^x(\tau - s) = 
\\
= \frac{1}{2\pi} \int_{-\infty}^{+\infty} d\omega\ \frac{J_x(\omega)\, e^{ \frac{\beta\omega}{2} }}{ \omega^2 \sinh\left( \frac{\beta\omega}{2} \right) }\, e^{ -i\omega\Delta t }
\end{multline}}

\end{itemize}

\section{Convergence tests\label{appdx:2baths}}
Figures~6 and 7 demonstrate numerical convergence of the dynamics of the TLS  with respect to memory time and time step $\Delta t$.

\begin{figure*}
\includegraphics[width=0.49\textwidth]{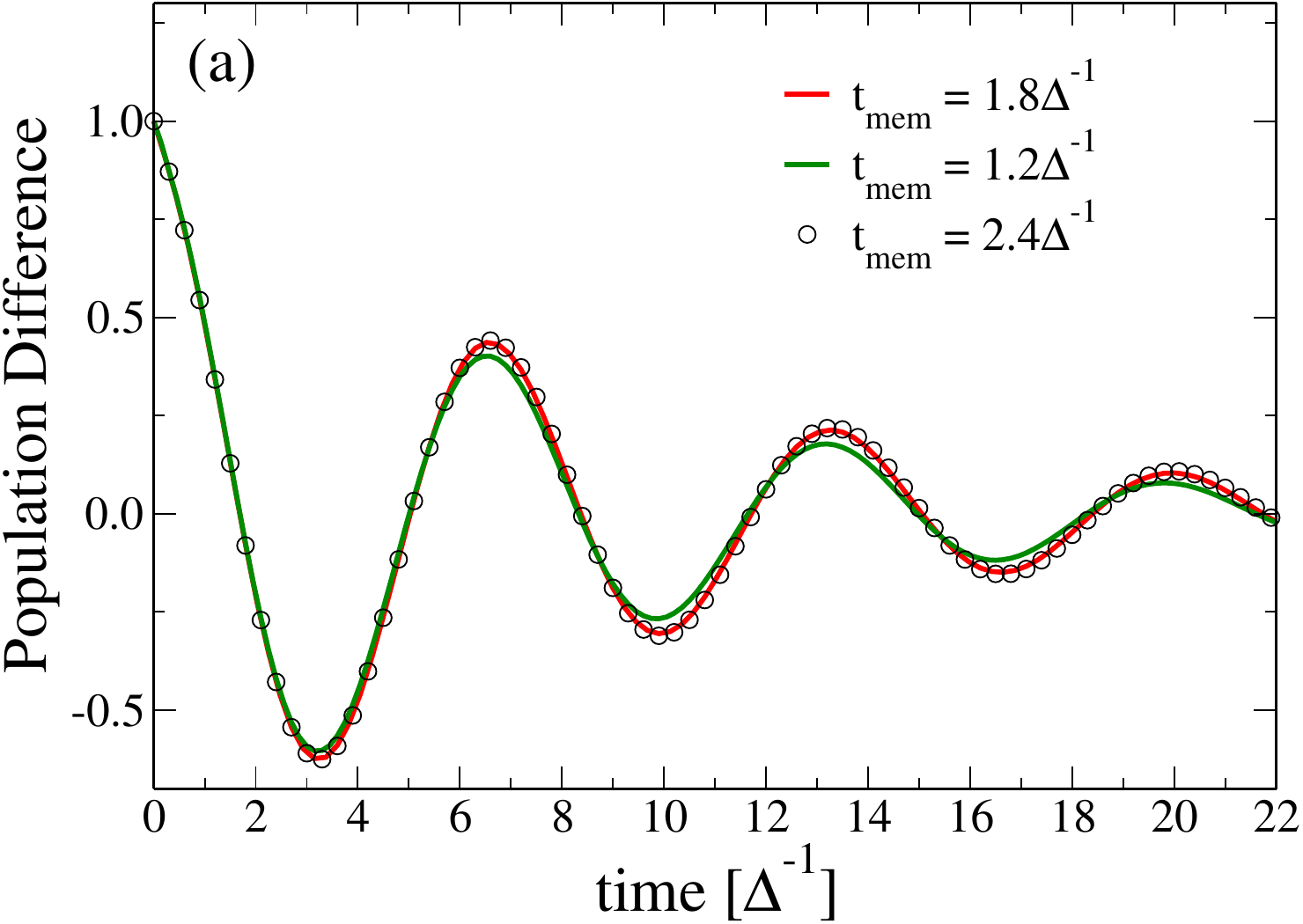}
\includegraphics[width=0.49\textwidth]{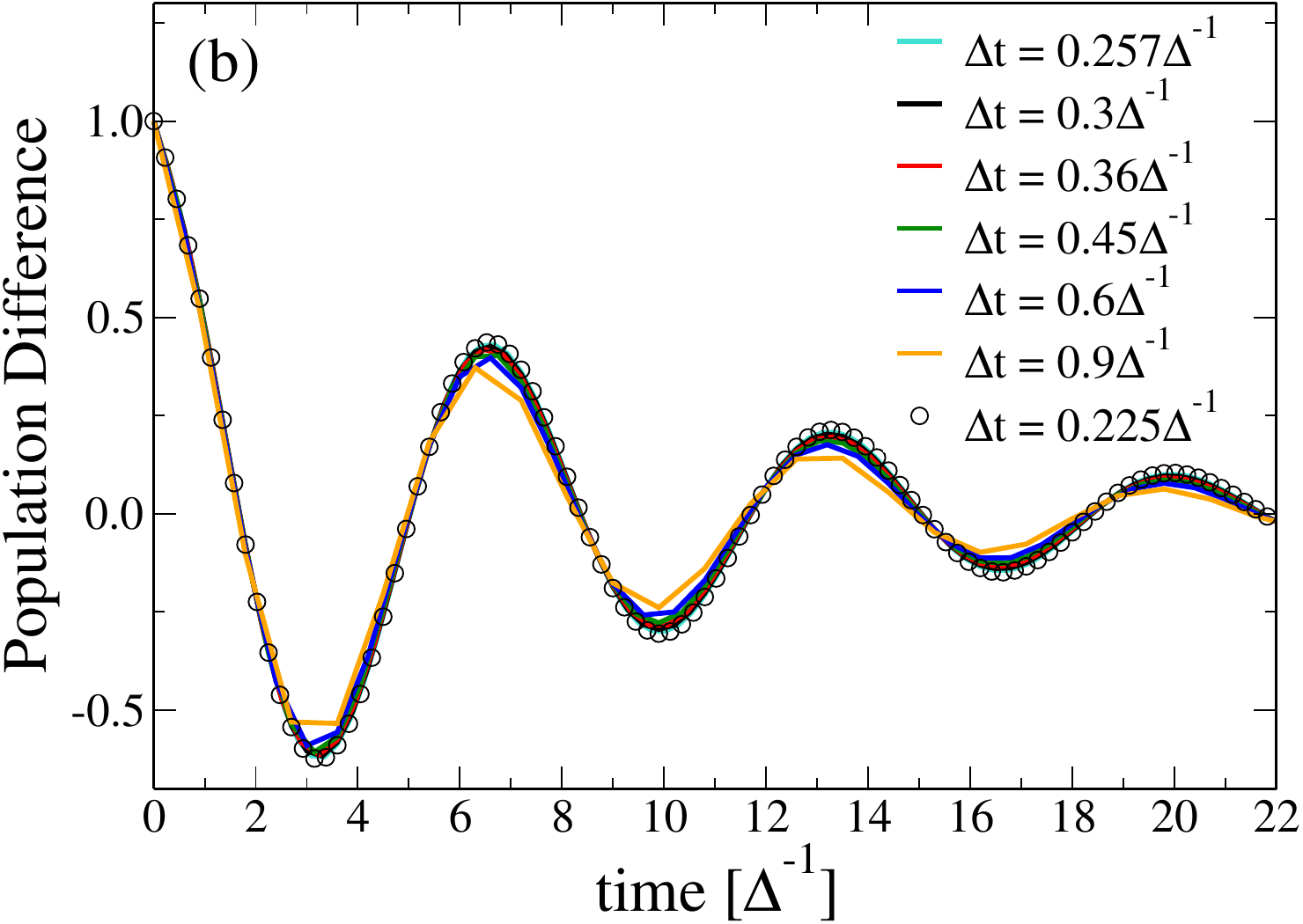}
\caption{
Convergence of system dynamics in the presence of two baths. On panel~(a), 
we fix the total number of paths by condition $N_\textrm{mask}^{x/z} = N_\textrm{mem}^{x/z} = 8$ and provide dynamics for three 
different memory cut-off and time steps: $t_\textrm{mem} = 1.8\Delta^{-1}$ and $\Delta t = 0.225\Delta^{-1}$ (red line), 
$t_\textrm{mem} = 1.2\Delta^{-1}$ and $\Delta t = 0.15\Delta^{-1}$ (green line), and 
$t_\textrm{mem} = 2.4\Delta^{-1}$ and $\Delta t = 0.3\Delta^{-1}$ (empty dots). On panel~(b), the memory time is fixed to 
the converged value of $t_\textrm{mem} = 1.8\Delta^{-1}$ and convergence with respect to different time steps $\Delta t$ is 
checked.
}
\end{figure*}
\begin{figure*}
\includegraphics[width=0.49\textwidth]{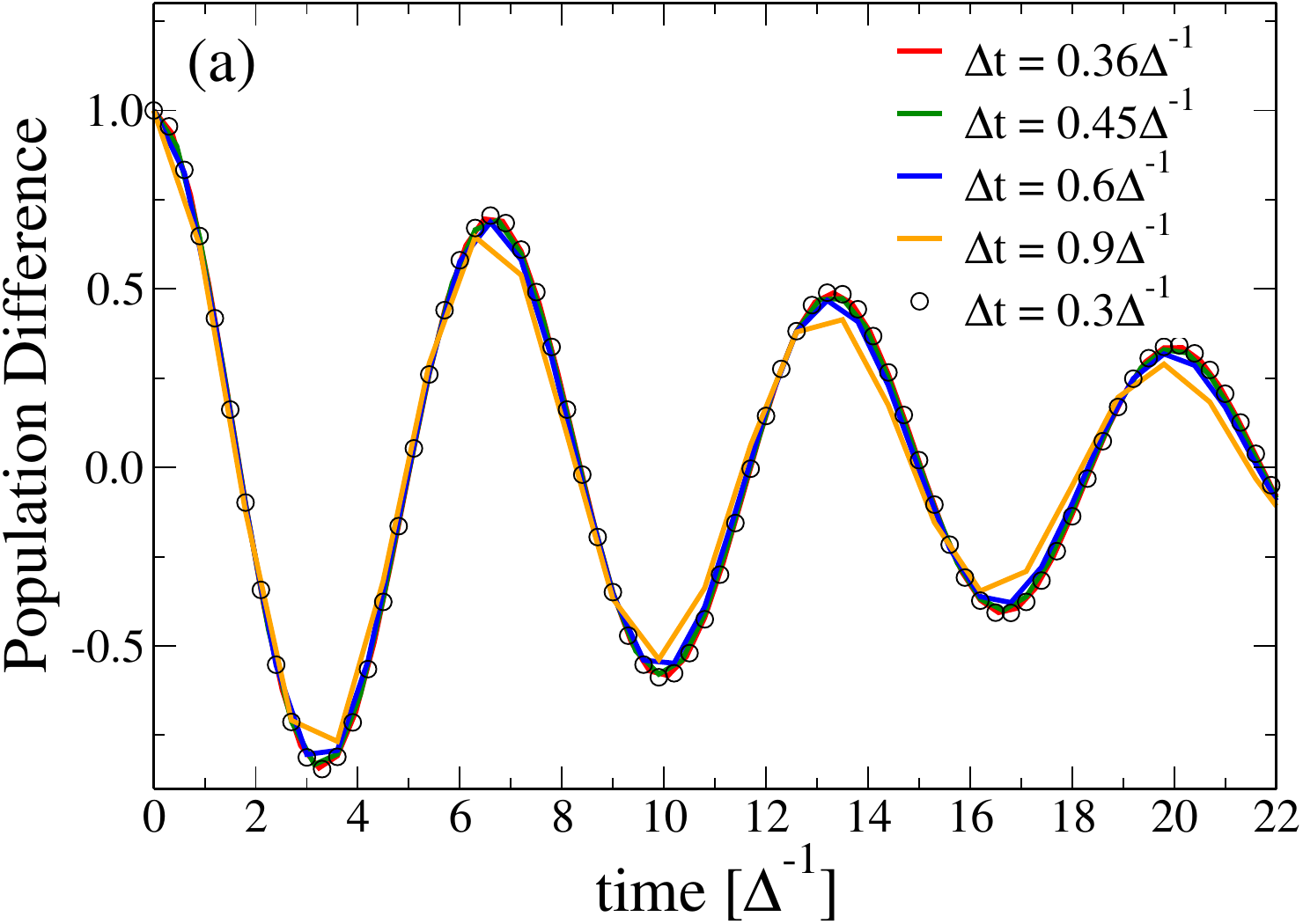}
\includegraphics[width=0.49\textwidth]{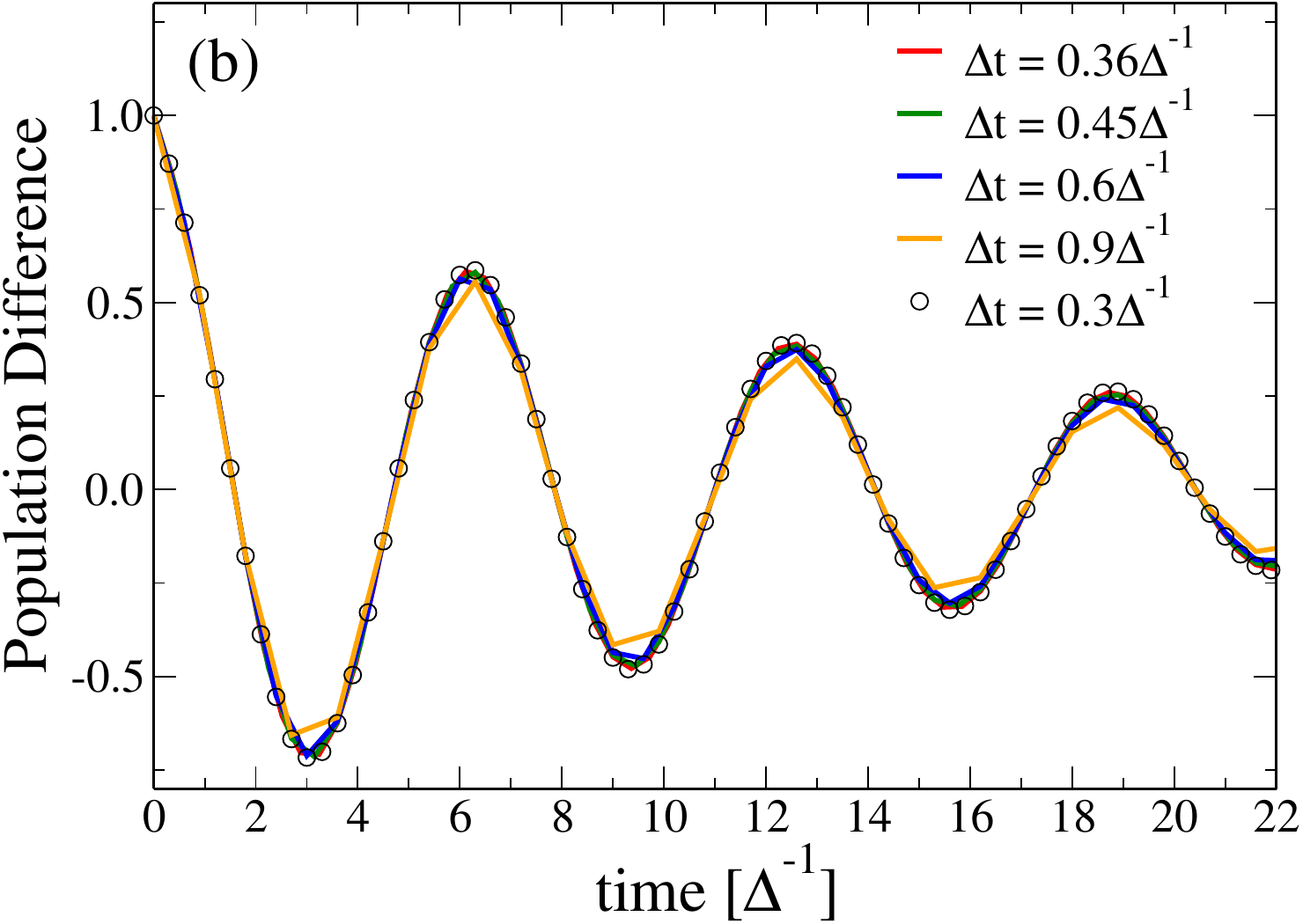}
\caption{
Convergence of the single general~(a) and single pure dephasing~(b) baths with respect to time step $\Delta t$. 
For panel~(a): $N_\textrm{mask}^\textrm{z} = N_\textrm{mem}^\textrm{z}$, and $t_\textrm{mem}^\textrm{z} = 1.8\Delta^{-1}$. 
For panel~(b): $N_\textrm{mask}^\textrm{x} = 2$ and $t_\textrm{mem}^\textrm{x} = 1.8\Delta^{-1}$. 
\label{fig:1Baths_conv_appdx}}
\end{figure*}

\FloatBarrier

\bibliography{manuscript}

\end{document}